\documentclass[review]{elsarticle}

\usepackage{comment}
\usepackage{mathtools}
\usepackage{lineno,hyperref}
\usepackage{multirow}
\usepackage{listings}
\usepackage{graphicx}
\usepackage{placeins}
\usepackage[ruled]{algorithm2e}

\usepackage{amssymb}
\usepackage{amsthm}
\usepackage{xcolor}
\usepackage{soul}
\definecolor{mygreen}{rgb}{0,0.6,0}
\definecolor{mygray}{rgb}{0.5,0.5,0.5}
\definecolor{mymauve}{rgb}{0.58,0,0.82}

\modulolinenumbers[5]

\journal{Journal of Microprocessors and Microsystems}

\begin{document}

\begin{frontmatter}

\title{The Multi-Dataflow Composer Tool: an open-source tool suite for Optimized Coarse-Grain Reconfigurable Hardware Accelerators and Platform Design}

%% Group authors per affiliation:
\author{Carlo Sau\textsuperscript{1}, Tiziana Fanni\textsuperscript{2}, Claudio Rubattu\textsuperscript{2},  Luigi Raffo\textsuperscript{1}, Francesca Palumbo\textsuperscript{2}}
\address{ \textsuperscript{1}University of Cagliari, \textsuperscript{2}University of Sassari}
%\fntext[myfootnote]{Since 1880.}

%% or include affiliations in footnotes:
%\author[mymainaddress,mysecondaryaddress]{Elsevier Inc}
%\ead[url]{www.elsevier.com}

%\author[mysecondaryaddress]{Global Customer Service\corref{mycorrespondingauthor}}
%\cortext[mycorrespondingauthor]{Corresponding author}
%\ead{support@elsevier.com}

%\address[mymainaddress]{1600 John F Kennedy Boulevard, Philadelphia}
%\address[mysecondaryaddress]{360 Park Avenue South, New York}

\begin{abstract}
\textcolor{black}{
Modern embedded and cyber-physical systems require every day more performance, power efficiency and flexibility, to execute several profiles and functionalities targeting the ever growing adaptivity needs and preserving execution efficiency. Such requirements pushed designers towards the adoption of heterogeneous and reconfigurable substrates, which development and management is not that straightforward. Despite acceleration and flexibility are desirable in many domains, the barrier of hardware deployment and operation is still there since specific advanced expertise and skills are needed. Related challenges are effectively tackled by leveraging on automation strategies that in some cases, as in the proposed work, exploit model-based approaches.}

\textcolor{black}{
This paper is focused on the Multi-Dataflow Composer (MDC) tool, that intends to solve issues related to design, optimization and operation of coarse-grain reconfigurable hardware accelerators and their easy adoption in modern heterogeneous substrates. MDC latest features and improvements are introduced in detail and have been assessed on the so far unexplored robotics application field. A multi-profile trajectory generator for a robotic arm is implemented over a Xilinx FPGA board to show in which cases coarse-grain reconfiguration can be applied and which can be the parameters and trade-offs MDC will allow users to play with.}
\end{abstract}


\end{frontmatter}

%\linenumbers

\section{Introduction}\label{s:intro}

Embedded system scenario has dramatically evolved in the last decades. Systems became ultra connected, bringing us to the era of Internet of Things. Then they started to massively interact with processes, humans and environment, becoming Cyber-Physical. Technologically speaking, there is not a standard template architecture for Cyber-Physical Systems (CPS), but designers are often requested to cope with complex evolving scenarios where multiple and distinct behavioral modalities have to be guaranteed. This implies that performance requirements may not be treated as fixed once and for all, and designers cannot base system design and deployments on always identical and predicable behaviors. CPS are required to be flexible to changeable functional (F) and non-functional (NF) requirements, being able to reconfigure their architecture and processing set-up according to environmental changes or unpredictable human requests, which implies the correct dynamic management of varying workloads and performance objectives.

\textcolor{black}{
Heterogeneity and adaptivity became then extremely appealing and desirable to address the challenging design and management of CPS. On  the  one hand, common single- or multi-core architectures are no longer capable to fulfill the demand for high efficiency, intended as resource, consumption and time in general, required by some applications, as the latest video coding standards, or by some execution contexts, like security or health related image and video processing.
At this purpose, alongside software cores, dedicated logic to process more efficiently such applications, when required, can be exploited in CPS, resulting thus in heterogeneous platforms. Nevertheless, enhanced capabilities do not come for free: dedicated hardware design requires specific skills, different from the common software ones.  
On the other hand, adaptivity can be granted by leveraging on reconfiguration. Nevertheless, while in software reconfiguration generally means programmability and it is easily supported, in hardware this is not the case since execution efficiency and flexibility are colliding requirements, which tend to further complicate both design and management of the heterogeneous substrate. 
Design and management effort is a well known issue of complex systems development, and can certainly be considered as the third major challenge in CPS deployment. Ideally, shortening time to market is a must to obey in the ICT world and it has historically been solved leveraging on models, to abstract away unnecessary details during the different design phases, and on design automation, which is favored by models and by the adoption of component-oriented design methodologies. }

\begin{figure*}
	\centering
	\includegraphics[width=1\textwidth] {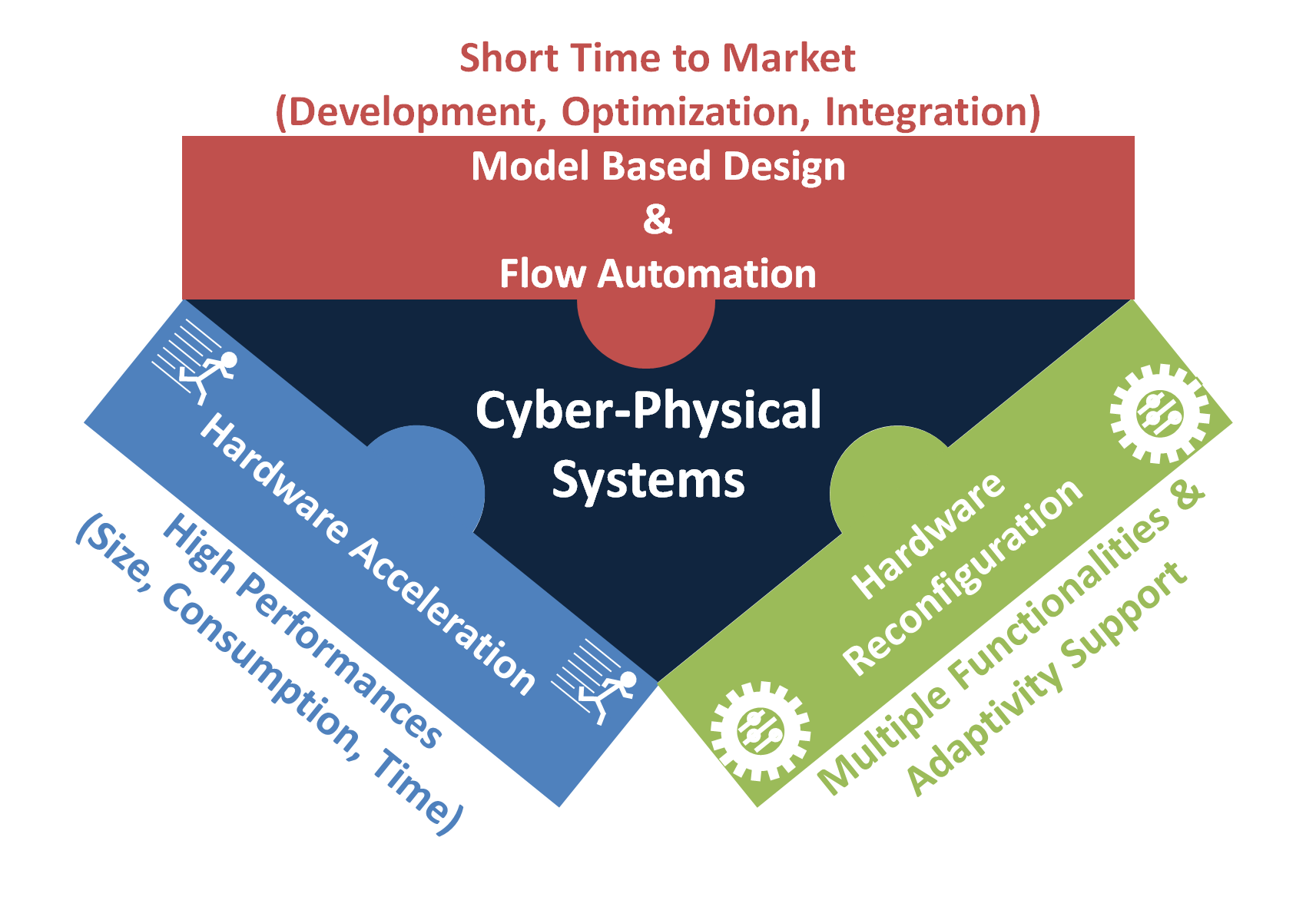}
	\caption{Designing Cyber-Physical Systems: possible solutions to known problems.}
	\label{fig:cps_problem}
\end{figure*}

\textcolor{black}{
Figure~\ref{fig:cps_problem} graphically summarizes the described CPS challenges and how they have been addressed in our studies: high performance by means of hardware acceleration, adaptivity and flexibility by means of hardware reconfiguration and, last by not least, time to market and minimization of required designer effort/skills by means of model-based design and programmability support for the proposed reconfigurable hardware accelerators. 
Our investments on defining a broad design environment for hardware accelerators is certainly aligned with the huge effort that big players on the market, as Xilinx, Intel, and Cadence, are dedicating in new instruments for hardware acceleration support \cite{Xilinx,Intel,Cadence}. On top of that, we have been focusing on reconfigurable hardware architectures since they have all the potentials to tackle the CPS adaptivity needs.} 

\textcolor{black}{In general, reconfigurable hardware architectures can be classified in different ways, one is per granularity.} \emph{Coarse-grain} reconfigurable architectures normally involve a fixed set of, often programmable, Processing Elements (PE) connected by means of dedicated routing blocks~\cite{Compton_2002}. These systems maximize resource re-use among different target applications by multiplexing in time the usage of PEs to serve different functionalities. \emph{Fine-grain} reconfigurable platforms are programmable at the single bit level. Field Programmable Gate Array (FPGA) platforms belong to this category and they became capable of changing context while executing, leveraging on Dynamic and Partial Reconfiguration (DPR) strategies~\cite{Altera:partial_reconf, Xilinx:partial_reconf}. Despite being appealing with respect to flexibility advantages, DPR has high costs in terms of \emph{1)} energy required to make the context switch, \emph{2)} memory usage required to store the configuration bitstreams, and \emph{3)} time to execute the switch. Different studies~\cite{PalumboFSRMDMPR19,PalumboSFR17,Yan_2012} demonstrated that adopting a Coarse-Grain Reconfigurable (CGR) approach can suite the scope of providing flexibility and adaptation to changeable F/NF requirements. 

For their high efficiency and lightweight reconfiguration capabilities, CGR systems are highly suitable bricks for the difficult to be built wall of a CPS, addressing the need for high performance and flexibility features, where different behaviors have to be supported.  
However, the main issue of CGR devices, totally in line with what we have seen for CPS development in general, is the complexity of their design under several aspects: resources mapping, optimization, hardware design, run-time management. This implies the urgency of developing automated methodologies for their design and management. The same world big players offering support for hardware acceleration do not provide straightforward support for CGR hardware solutions. In literature, this lack has been addressed by research works proposing automated and semi-automated strategies \cite{Palumbo_2016, Ansaloni_2012}. 
In particular, in \cite{Sau_2015} the Multi-Dataflow Composer (MDC) has been introduced to provide design automation for CGR accelerators. Accelerators, basically custom IPs for Xilinx environments, are automatically derived  from dataflow models. Limitations in terms of system integration capabilities were still experienced in \cite{Sau_2015} and have been overcome by the present work, as detailed below. 

\subsection{Contribution of the Work}
\textcolor{black}{The contributions of this paper, and more in general what is new in the MDC tool with respect to \cite{PalumboFSRMDMPR19}, are reported hereinafter. 
\begin{enumerate}
	\item The tool and its baseline and advanced features for the first time are available open source, and are presented all together in a comprehensive manner.
	\item The coprocessor generator, MDC system integration advanced feature, has been made fully compatible to Xilinx Vivado Design Suite to provide straightforward system integration, and it has been substantially enhanced:
	\begin{itemize}
    	\item it is now supporting a largely used host hard-core, the ARM, along with the previously supported Xilinx proprietary soft-core, the MicroBlaze;
    	\item it is now supporting a largely used system bus, the ARM AMBA AXI4, drastically increasing the number of compatible potential target platforms, and letting it possible to provide more efficient coprocessor interfaces by exploiting different available bus protocols according to the nature of the transmission (control or data);
    	\item it is now possible to adopt a Direct Memory Access (DMA) engine to relieve the host processor from the burden of taking care about data transfers;
    	\item it is now providing TCL scripts to have faster and simpler system integration, delivering real ready-to-use CGR hardware accelerators.
    \end{itemize}
	\item The tool has been assessed on a completely new application scenario, robotics that is far away from the usual image/video processing scenarios where MDC proved its capabilities and potentials before.
	\item The model-based approach followed within MDC made it suitable for the integration with other tools, resulting in a more complete and powerful CPS design and management support.
\end{enumerate}
}

\subsection{Structure of the Work} 
\textcolor{black}{
The rest of the paper is organized as follows.
\begin{itemize}
    \item Section~\ref{s:back} describes the background of the MDC tool and of its features, from baseline to advanced ones, providing a brief state of the art for each of them. 
    \item Section~\ref{s:baselineMDC} gives an overview of the MDC tool and of its features, with emphasis on the new functionalities and improvements introduced in this work.
    \item Section~\ref{s:ass} provides the assessment of the MDC tool, and in particular of the new functionalities and improvements, within the robotics application field.
    \item Section~\ref{s:usage} reports an evaluation of MDC tool in terms of usability and design effort.
    \item Section~\ref{sect:interf} shows the enhanced possibilities enabled by the interoperation of MDC with other existing tools in the context of heterogeneous and adaptive CPS.
    \item Section~\ref{s:concl} concludes the paper with some final remarks and future directions.
\end{itemize}}

\section{Background}\label{s:back}

\textcolor{black}{Reconfigurable computing refers to a class of digital electronic system architectures that combine the flexibility typical of software programmed systems to the high performance of the hardware implementations~\cite{Compton_2002}.
Reconfigurable systems are often called adaptive, meaning that the logic units and interconnects of the system can be modeled to fit a specific functionality by programming it at  hardware level~\cite{Tessier_2001}. However, the more these components are able to fit the applications requirements, the slower they are with respect to less flexible component, which can easily turn out to be also smaller in area and less power consuming~\cite{Todman_2005}.
}
\textcolor{black}{As already said in Section \ref{s:intro}, CGR systems provide word-level reconfigurability and, despite being customizable over a smaller number of scenarios, they reconfigure faster than fine-grain ones.
}

\begin{figure*}
	\centering
	\includegraphics[width=0.7\textwidth] {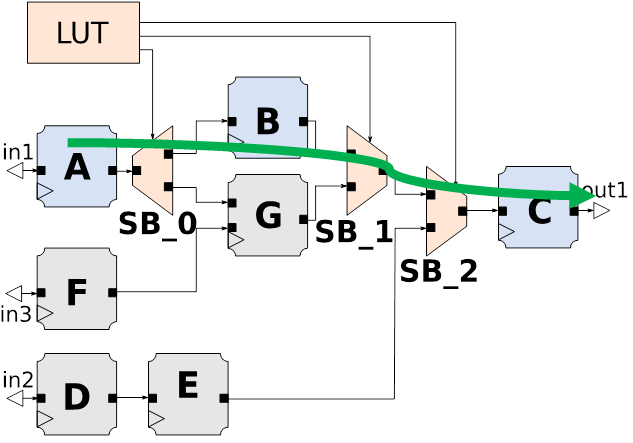}
	\caption{Example of a Coarse-Grain Reconfigurable Circuit.}
	\label{fig:cgr_circuit}
\end{figure*}

\begin{figure*}
	\centering
	\includegraphics[width=1\textwidth] {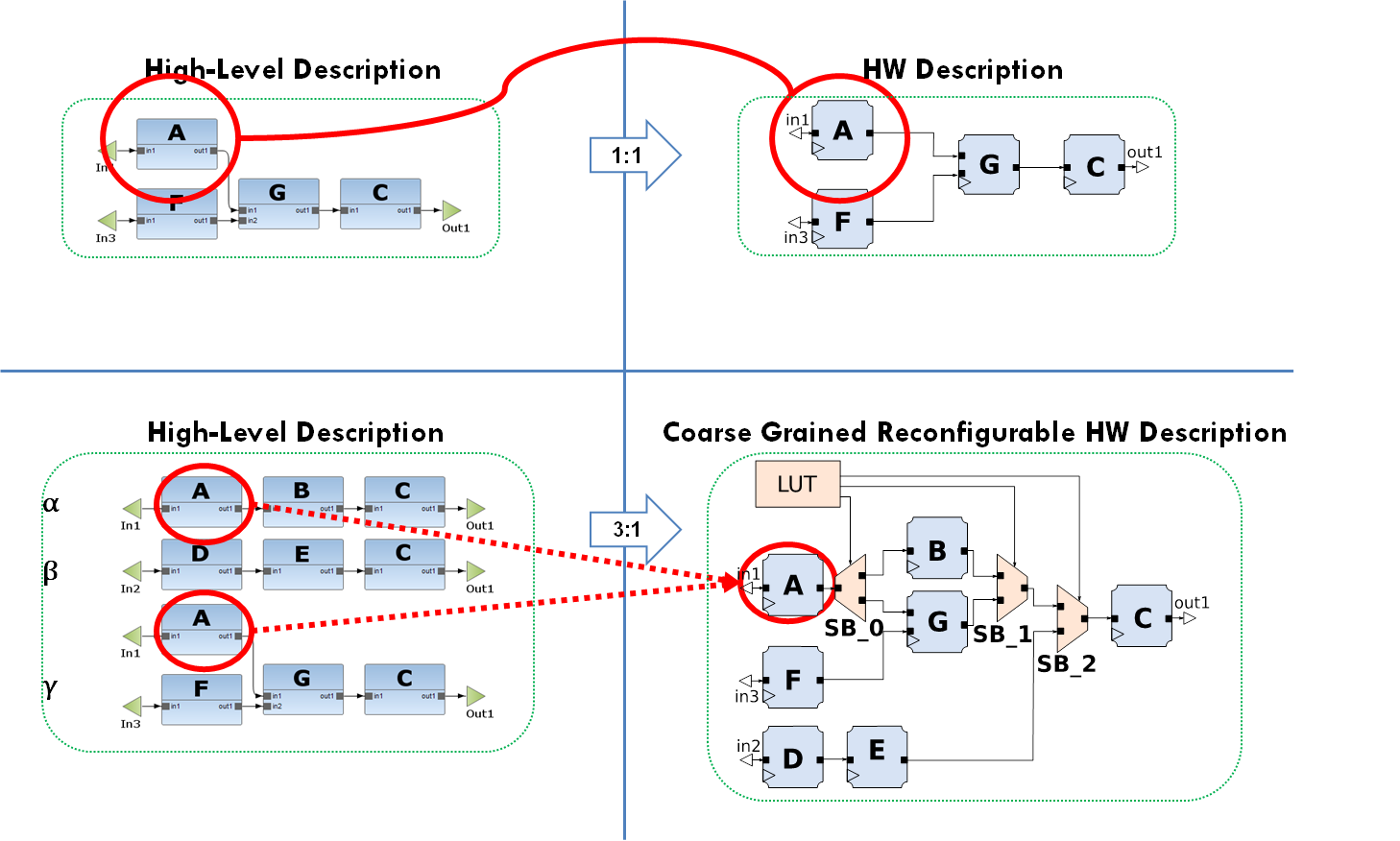}
	\caption{High-level of abstraction for hardware description.}
	\label{fig:dataflow2hw}
\end{figure*}

\textcolor{black}{
In this paper we focus on heterogeneous CGR systems and, more precisely, on systems able to compute different functionalities, decided at the design time. All the necessary logic is deployed on the computing substrate at design time, and common resources are shared, but only one functionality per time can be enabled. Such kind of accelerators is suitable to be deployed both on FPGAs and on Application Specific Integrated Circuits (ASICs). Figure \ref{fig:cgr_circuit} illustrates an example of a CGR circuit, able to execute two different functionalities. When the first functionality is enabled, PEs \emph{A}, \emph{B} and \emph{C} are activated through the proper setting of the multiplexers, placed at the crossroads of the paths, while the remaining logic is in an idle state. \textcolor{black}{Please note that this does not mean that it is disabled, but only that it is not involved in the current computation.}} \textcolor{black}{
The more are the functionalities to be implemented in the CGR circuit, the more is the design complexity. Indeed, identifying the logic that can be shared among the functionalities, to minimize both the number of resources and their connections, and properly managing activation paths at run-time are not straightforward.  Generally speaking, it is possible to model a hardware representation with a higher-level of abstraction and to transform the model into a circuit by means of a 1:1 mapping process (see top part of Figure \ref{fig:dataflow2hw}). However, this transformation and mapping process, when the number of functionalities increases and optimizations have to be applied, gets complex and requires automation.
}

\textcolor{black}{The rest of this section is organized as follows: 
\begin{itemize}
    \item the dataflow Models of Computation (MoCs) are described in Section \ref{Soa:dataflow} to understand which kind of inputs MDC users have to master, along with the features and characteristics that make dataflows appropriate to solve CGR related design problems; 
    \item the works in literature addressing design issues in reconfigurable and digital signal processing contexts exploiting dataflow MoCs are discussed in Section \ref{ss:tool_dataflow}, in order to position MDC in the plethora of available dataflow-based state of the art tools for digital design;
    \item the power issue that CGR systems, and digital circuits in general, have to face is addressed in Section \ref{Soa:power}. Please note that power/energy is one of the most important metrics in the design and run-time management of CPS. Therefore, despite not being among the goals of this paper, for the sake of completeness, the power related MDC feature and its scientific roots were worth to be introduced.
\end{itemize}
}

\subsection{Dataflow-Based Design}
\label{Soa:dataflow}

\begin{figure*}
	\centering
	\includegraphics[width=4in] {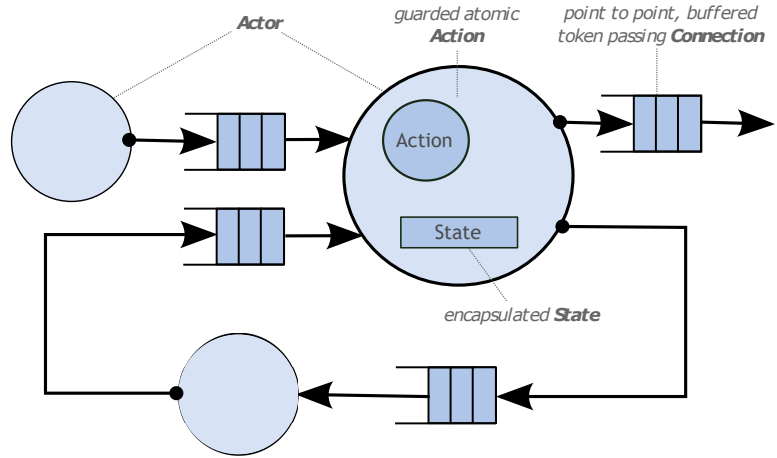}
	\caption{Dataflow Process Network design example.}
	\label{fig:dpn}
\end{figure*}

\textcolor{black}{
Model-based design has been widely studied and applied over the years in many domains of embedded processing. Dataflow is well-known as a paradigm for model-based design that is effective for embedded digital signal processing (DSP) systems~\cite{Bhat_2013x1, Dennis_1974, Kahn_1974}. 
A dataflow can be described as a direct Data-Flow Graph (DFG) $DFG\langle V,E\rangle$, where $V$ is the set of vertices of the graph (the actors) and $E$ is the set of edges representing loss-less, order-preserving point-to-point connection channels. One of the first formalizations of dataflow models has been presented by Lee et al.~\cite{Lee_1995} with the Dataflow Process Networks (DPNs) illustrated in Figure~\ref{fig:dpn}. The \emph{actors} are abstract representations of PEs that encapsulate their own internal state and asynchronously concur to the whole computation. The communication  between actors is based on the exchange of sequences of atomic data packets called \emph{tokens}. This communication is asynchronous, since it is driven by the production and consumption of tokens. Once triggered for processing (\emph{fired}), actors execute a sequence of steps called \emph{actions} that can result in: (1) the consumption of one or more input tokens; (2) the production of one or more output tokens; (3) the change of the actor internal state.}

\textcolor{black}{
This model is suitable to manage the concurrency due to parallelism that one application may intrinsically have. Indeed, thanks to the token mediated communication policy, race conditions among actors are avoided. Furthermore, dataflows are highly modular specifications naturally amenable to block diagrams; therefore, perfectly fitting to signal processing applications.
Actors can be implemented by any host language able to specify the actions firing rules. This includes the possibility to specify them also as Intellectual Properties (IPs) coded in Hardware Description Language (HDL), low-level software actors written in C and high-level software actors written in Java.  Modularity strongly favors the code reuse, speeding-up the time to market needed for updating sub-parts of already existing applications or for modeling new functionalities from scratch. All these distinctive features make dataflows very suitable for programming highly parallel, also heterogeneous systems, like multi-processor systems on chip or CGR arrays.}

\subsection{DSP-oriented Dataflow-based Tools}
\label{ss:tool_dataflow}
\textcolor{black}{
Due to their distinctive features, dataflow models are adopted in a wide variety of tools for both software and hardware design.  
% dataflow to hardware
In~\cite{McAllister_2004}, methodologies for modeling, implementing and optimizing 
pipelined hardware component networks from a high-level dataflow graph description have been developed. They offer the possibility of optimizing the design in terms of throughput or resource consumption. 
% COMPAAN and LAURA
Stefanov et al.~\cite{Stefanov_2004} present a system design flow, centered around the exploitation of the Kahn Process Network model, in which an application written in a subset of MATLAB is mapped onto a target platform composed of a Central Processing Unit (CPU) and an FPGA in a systematic and automated way. To realize the flow, they developed and used the COMPAAN and LAURA tools, to go from an application specification in MATLAB to an implementation of the application running on the target platform.
%preesm
PREESM is an open-source Eclipse-based tool that provides dataflow-based methods to efficiently run applications on a multicore DSP system~\cite{Pelcat_2014}. PREESM provides the designer with information on algorithm parallelism and latency estimates, as well as on system memory requirements. It automatically maps and schedules the application, specified as Parameterized and Interfaced Synchronous Dataflow (PiSDF) MoC \cite{Desnos_2013}, over the available PEs, and provides a code generation feature to transform the dataflow representation into a compilable code.}

\textcolor{black}{
% ORCC and CAL
A substantial work on dataflow-based tools regards the MPEG Reconfigurable Video Coding (MPEG-RVC), an initiative to enhance video codecs interoperability through the adoption of dataflow models and of language similar to C for describing actors, the Caltrop Actor Language (CAL). Most of the tools around MPEG-RVC leverage on the Open RVC-CAL Compiler (ORCC)~\cite{Orcc}, a compilation infrastructure in charge of generating descriptions in several languages (software, hardware or mixed for co-design~\cite{Siret_2010}) starting from CAL actors and XML Dataflow Format (XDF) networks. At the moment ORCC is provided as an Eclipse plug-in written in Java and relies on an Intermediate Representation (IR) of the DPNs that is still specified in Java. The IR can be exploited to feed several other tools such as Turnus~\cite{Brunet_2013}, which offers simulation, profiling and design space exploration capabilities, and Xronos~\cite{Bezati_2013}, in charge of providing C code ready to be used in HLS for Xilinx FPGAs. The ORCC compilation infrastructure, as well as the MPEG-RVC framework itself, is continuously evolving in order to support new and more advanced features and to produce more and more dynamic systems.
}

\textcolor{black}{
The CAPH language and related framework represent another recent effort to generate HDL from a dataflow language~\cite{Serot_2013,Serot_2016}. More precisely CAPH is a toolchain built around a domain-specific language for the specification of stream-processing applications based on a dynamic dataflow MoC. This latter is specified through a functional language named Functional Graph Notation (FGN)~\cite{Serot_2008}, allowing a complete description of a dataflow network by means of purely functional expressions, and resulting in improved abstraction capabilities, easier wiring description and more efficient errors check.}

\textcolor{black}{
The Lightweight dataflow (LWDF) is a programming methodology that allows designers to
systematically integrate and experiment with dataflow modeling approaches in the context of existing design processes~\cite{Shen_2010}. LWDF is ``lightweight'' in the sense that the programming model is designed to be minimally intrusive on existing design methodologies and processes. It delivers a compact set of Application Program Interfaces (APIs) that can be used to incorporate advanced dataflow techniques and requires minimal dependence on specialized tools or libraries.
}

\textcolor{black}{In none of the literature works dataflows have been used to address CGR systems development, optimization and management. The only works on this topic are related to the MDC tool itself that is the object of the proposed work.}

\subsection{The Power Issue}
\label{Soa:power}
\textcolor{black}{
As said at the beginning of this section, CGR systems execute different functionalities, multiplexing resources in time. The logic that is not involved in the currently running computation is in an idle state and, necessarily, uselessly consuming power. In digital systems, power consumption can be divided onto two main contributions: static and dynamic (see Equation~\ref{eq:pow_tot}). The former is always present when the circuit is powered on, since it is due to leakage currents ($P_{lkg}$). The latter is dissipated only when logic transitions occur, so that it is related to the switching activity during the system execution. }

%%%%%%%%%%%%%%%%%%%%%%%%%%%%%%%%%%%%%%%
\begin{equation}\label{eq:pow_tot}
\begin{split}
P_{tot} = P_{static} + P_{dynamic}
\end{split}\end{equation}
%%%%%%%%%%%%%%%%%%%%%%%%%%%%%%%%%%%%%%%

\textcolor{black}{
Several techniques (clock gating, multi-frequency, operand isolation, multi-threshold, multi-supply libraries, power gating, etc.) can be applied to reduce power consumption and, in some cases, they are automatically implemented by commercial synthesis/place-and-route tools.}

\textcolor{black}{
Clock gating is a really popular technique that consists on shutting off the clock of the unused synchronous logic, reducing the dynamic power consumption due to the clock tree and to sequential logic up to the 40\%~\cite{Zhang_2006}. Clock gating has been deeply employed for more than 20 years~\cite{Pedram_1996,Wu_2000}. Commercial synthesizer such as Cadence  RTL Compiler (or the more recent Genus)~\cite{Cadence:RTL,Cadence:Genus}, or Synopsys Design Compiler~\cite{Synopsys:DC} are able of automatically gating groups of flip-flops when enabled by the same control signal. At the state of the art, some works focused instead on the application of clock gating at a higher-level, targeting FPGAs ~\cite{OZBALTAN_2018, Bezati_2017}. In particular Bezati et al.~\cite{Bezati_2017} presented an extension of a dataflow-based High-Level Synthesis (HLS) tool, Xronos, to selectively switch off clock signal for parts of the circuit that are idle due to stalls in the pipeline.  }

\textcolor{black}{
More complex power saving strategies such as, voltage/frequency scaling~\cite{Herbert_2007,Eyerman_2011} and power shut-off schemes~\cite{Arora_2014,Jeff_2012}  can be extremely beneficial.
Nowadays, some of the electronic design automation companies offer support for automatically integrating low power techniques, such as clock gating, dynamic voltage/frequency scaling or power gating, but this is mainly a specification support and most of the job is still manually done by designers who have to define the power format file ~\cite{IEEE:UPF, SI2CPFspecification}.
This process can be error prone and time consuming, and also not easily applicable to automatically generated CGR systems, as the ones considered in this paper.
}

\textcolor{black}{
Recently some works focused on the application of power saving methodologies, automatically generating a power format file.
In \cite{Gagarski_2016} authors present a SCPower extension that allows to inject power specification into synthesizable hardware designs in SystemC language, providing the automatic generation of the Unified Power Format (UPF) file, compatible with the Synopsys environment. However, this work focuses more on enabling power-aware verification of SystemC designs.
Qamar et al. \cite{Qamar_2016} present a methodology that considers the application of clock and power gating techniques to the register transfer level (RTL) systems generated automatically by HLS, using SystemC code. At high-level of abstraction, they specify the power intent, to generate the Common Power Format (CPF) file, compatible with Cadence tools, to implement the power gating. However, this work still requires hand-work. Indeed, it mainly moves the definition of the power intent from RTL level to higher-level, specifying it through the insertion of pragma into the SystemC code. Furthermore, the logic to be switched off through power saving techniques is not automatically identified. To automate power-management specification, Macko \cite{Macko_2018} proposed another method, which requires as input a system functional model in SystemC and electronic system level simulation results. The output is an enriched system model, which includes the power-management specification using SystemC/PMS. However, this method is limited to SystemC high-level description, and is not applicable to CGR systems.}

\section{The Multi-Dataflow Composer Tool}
\label{s:baselineMDC}

This section describes the Multi-Dataflow Composer \footnote{Available on GitHub: https://github.com/mdc-suite/mdc} that, as already said, is an open-source automated tool for the generation and management of Coarse-Grain Reconfigurable (CGR) multi-functional architectures.
MDC is meant to address the difficulty of mapping a set of different applications onto a CGR architecture~\cite{Carta_2006,Kumar_2006}, combining together a set of input dataflow specifications describing the desired system behaviors. MDC is capable of identifying the actors that can be shared among the input dataflow specifications and applies a datapath-merging problem-solving algorithm to generate a CGR hardware substrate \cite{Palumbo_2016}. \textcolor{black}{The baseline MDC approach is target and technology independent, indeed the CGR circuits it generates can be implemented on FPGA or ASIC, with any tool for digital design. However, some of MDC features are target or technology dependent.}
Figure \ref{fig:mdc-components} illustrates the four main MDC components:  

\begin{figure*}[h!t]
	\centering
	\includegraphics[width=0.9\textwidth]{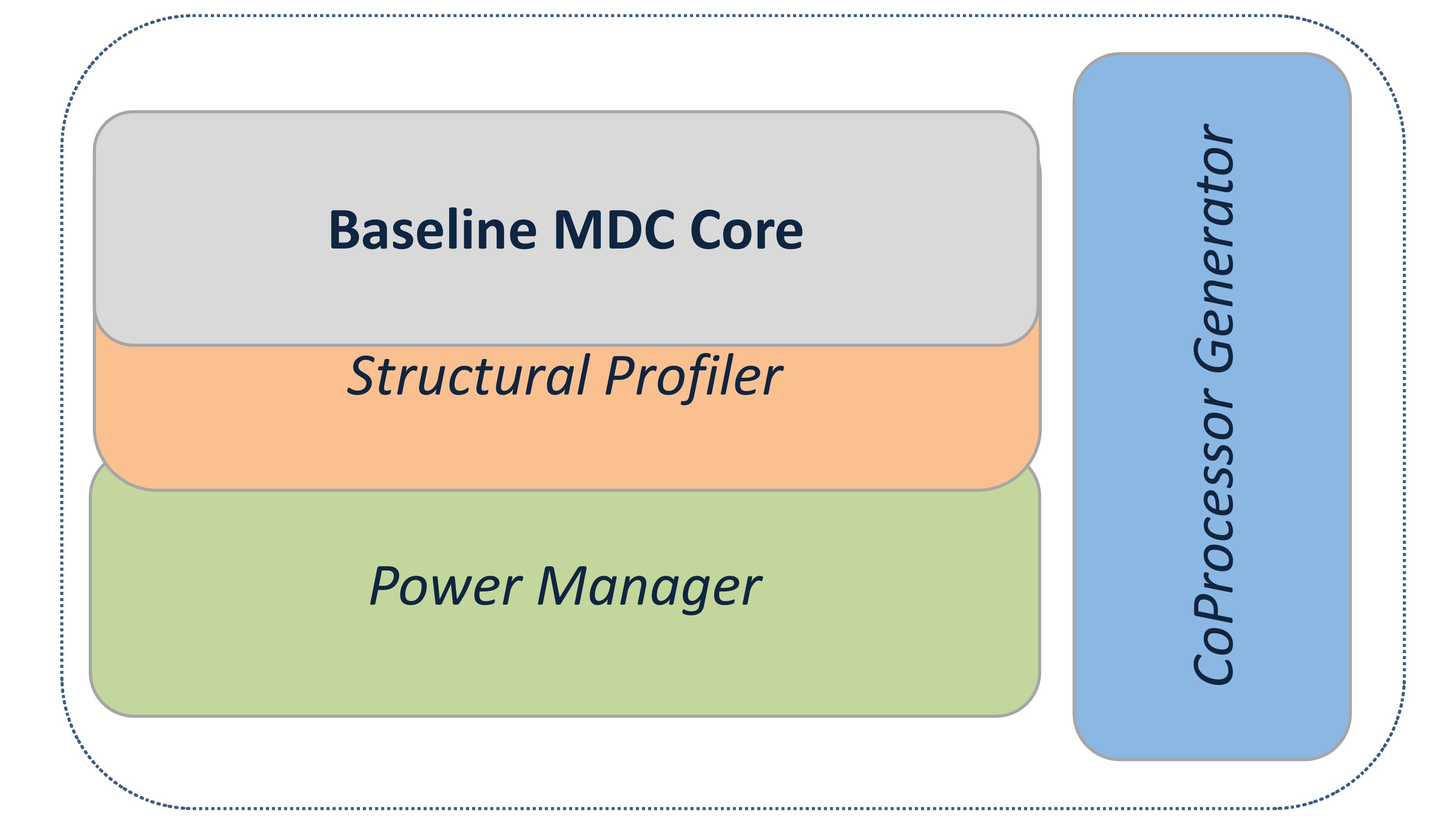}
	\caption{Multi-Dataflow Composer tool components}
	\label{fig:mdc-components}
\end{figure*}

\begin{itemize}
	\item \textit{Baseline MDC Core - (Section \ref{ss:baselineMDC})}: performing dataflow-to-hardware composition, by means of datapath merging techniques \cite{Souza_2005}. \textcolor{black}{The baseline MDC core generates a CGR datapath that can be implemented on both FPGA or ASIC.%, without any limitation on the digital design tool adopted.
	}
	\item \textit{Structural Profiler - (Section \ref{ss:profilerMDC})}: performing the design space exploration of the implementable
	multi-functional systems, which can be derived from the input dataflow specifications set,
	to determine the optimal CGR substrate according to the given input
	constraints \textcolor{black}{\cite{Palumbo_2016}}. %\cite{Palumbo_2014x2}. 
	\textcolor{black}{This feature is available for ASIC implementations only at the moment.}
	\item \textit{Dynamic Power Manager - (Section \ref{ss:dynPowerMDC})}: performing, at the dataflow level, the logic partitioning of the
	substrate to implement at the hardware level a clock gating or power gating strategy, and system modelling \cite{Palumbo_2016}. \textcolor{black}{The MDC power saving can be applied to both FPGA or ASIC, when the clock gating is chosen, while it can be applied only to ASIC when the power gating is involved.}
	\item \textit{Coprocessor Generator - (Section \ref{ss:coprMDC})}: performing the complete dataflow-to-hardware customization of a Xilinx compliant multi-functional accelerator that can be either loosely coupled or tightly coupled to the main processor, according to the processing needs. Drivers and scripts for fast system integration are also automatically derived. %Starting from the input dataflow specifications set, the customized accelerator can be either loosely coupled or tightly coupled to the main processor, according to the design needs, and also its drivers and script for fast system integration are derived 
	\textcolor{black}{This feature will be deeply discussed in the related section being one of the main contributions of this work extending what was presented in \cite{Sau_2015}}.
\end{itemize}

\subsection{Baseline MDC Core}
\label{ss:baselineMDC}

The core functionality of MDC tool is in charge of mapping a set of dataflow specifications onto
a CGR substrate, automating the mapping process while minimizing hardware 
resources. This issue is known in literature as the datapath merging 
problem \cite{Souza_2005}. MDC solves it by exploiting two different \textcolor{black}{iterative merging algorithms}: (1) a heuristic
algorithm \cite{Palumbo_2016}, or (2) Moreano's algorithm~\cite{Moreano_2002}.

The tool is designed to be connected to higher-level utilities by means of an adequate
front-end, in charge of parsing the high-level descriptions of the datapaths to be combined.
In this way, relying on the chosen front-end, MDC is able to process any type of DFG. %\footnote{$DFG\langle V,E\rangle$ is a directed graph, where $V$ is the set of vertices of the graph (the actors) and $E$ is the set of edges.}. 
MDC has been coupled with different dataflow-based tools, such as ORCC~\cite{Orcc}, CAPH~\cite{Serot_2013} and Synflow~\cite{Synflow}. In this manuscript, the coupling between ORCC and MDC, and the DPNs, expressed as XDF files, are used to illustrate MDC features.

\begin{figure*}[h!t]
	\centering
	\includegraphics[width=0.9\textwidth]{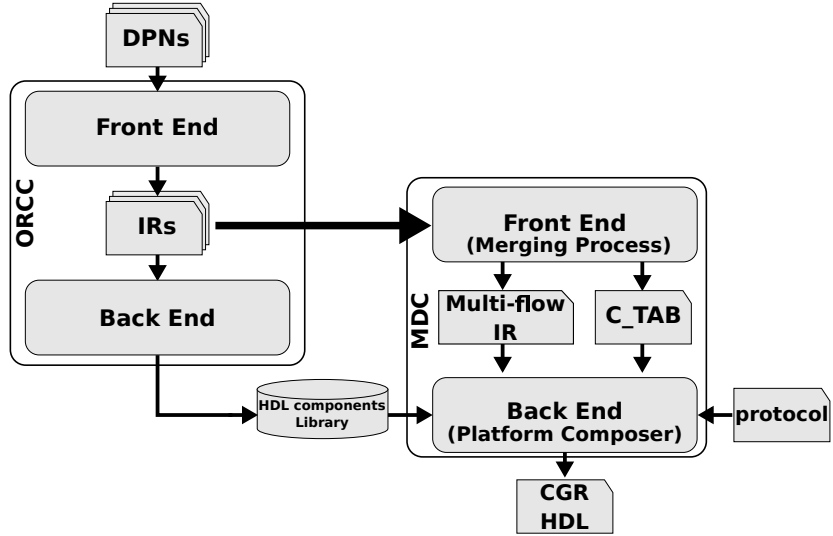}
	\caption{Multi-Dataflow Composer tool: an overview.}
	\label{fig:orccFlow}
\end{figure*}

Figure~\ref{fig:orccFlow} shows an overview of the coupled ORCC-MDC 
design flow. Starting from the DPN models of the functionalities to be implemented,
three major steps are required to generate the HDL specification of a
multi-functional reconfigurable datapath: 1) input DPNs parsing; 2) multi-dataflow generation; and 3) generation of the HDL description of the CGR hardware architecture. 

ORCC parses the input DPNs, along with their actors, and translates each of them into a DFG Java Intermediate Representation (IR). During the parsing, ORCC explodes non-atomic actors (composed
of a sub-network of actors), flattening the input DPNs. \textcolor{black}{As depicted in Figure~\ref{fig:orccFlow}, ORCC provide several IRs, one for each input DPN.}
Then the MDC front-end leverages on the IR translations to assemble a single multi-functional dataflow network
(\emph{Multi-flow IR} in Figure~\ref{fig:orccFlow}). During this phase, MDC front-end keeps trace of the
system programmability through the \emph{Configuration Table}  (\emph{C\_TAB} in
Figure~\ref{fig:orccFlow}). 
Reconfiguration is implemented by multiplexing resources in time. Ad-hoc low overhead switching modules (Switching Boxes - SBoxes) are placed at the crossroads between the different paths of data and driven by dedicated Look-Up Tables (LUTs), whose content is defined according to the \emph{Configuration Table}.  Once the input DPNs have been merged, the MDC back-end creates the hardware description (\emph{CGR HDL} in Figure~\ref{fig:orccFlow}), mapping each actor onto a  different PE. Even though MDC is coupled with ORCC, the generated CGR hardware is not restricted to the RVC-CAL communication protocol. Indeed, MDC takes as input an XML file
that describes communication protocol between PEs (\emph{protocol} in Figure~\ref{fig:orccFlow}). Thus, MDC is actually able of considering 
a dataflow network as generic graph, where communication among PEs can be managed
with or without First-In First-Out (FIFO) connections, and where the PEs can even be purely
combinatorial. The HDL description of the PEs are passed as input to MDC, together with 
any other necessary module (e.g. FIFOs, fanouts, memories, etc.) within the \emph{HDL components library} 
(see Figure~\ref{fig:orccFlow}) that can be manually written or automatically created by HLS tools. In the tool flow shown in 
Figure~\ref{fig:orccFlow}, the \emph{HDL component library} is created by an ORCC backend. \textcolor{black}{Please note that the figure reports on the original MDC baseline core composition. Currently the ORCC backend for HDL code generation has been dismissed. The \emph{HDL components library}, when automatically generated, is normally created in our designs either with CAPH or Vivado HLS.} 

In the current HDL implementation, SBoxes are combinatorial multiplexers; therefore, no 
dedicated FIFO buffers are inserted with the SBox units. Nevertheless, the FIFOs of the
upstream/downstream actors have to be managed. $Sbox\_1\times2$ units, inserted to split a 
path of data, require one FIFO for each outgoing connection. In the case of $Sbox\_2\times1$ units, inserted to access a common shared actor, the FIFO buffers are placed before the SBox along the
incoming connections. Since the SBoxes are fully combinatorial and the FIFO buffers always
belong to the other actors, the well known dataflow problem of the FIFO buffers optimal sizing
does not affect the MDC merging process. Input DPNs have only to be properly sized before the
MDC execution.

\begin{figure*}[h!t]
	\centering
	\includegraphics[width=1\textwidth]{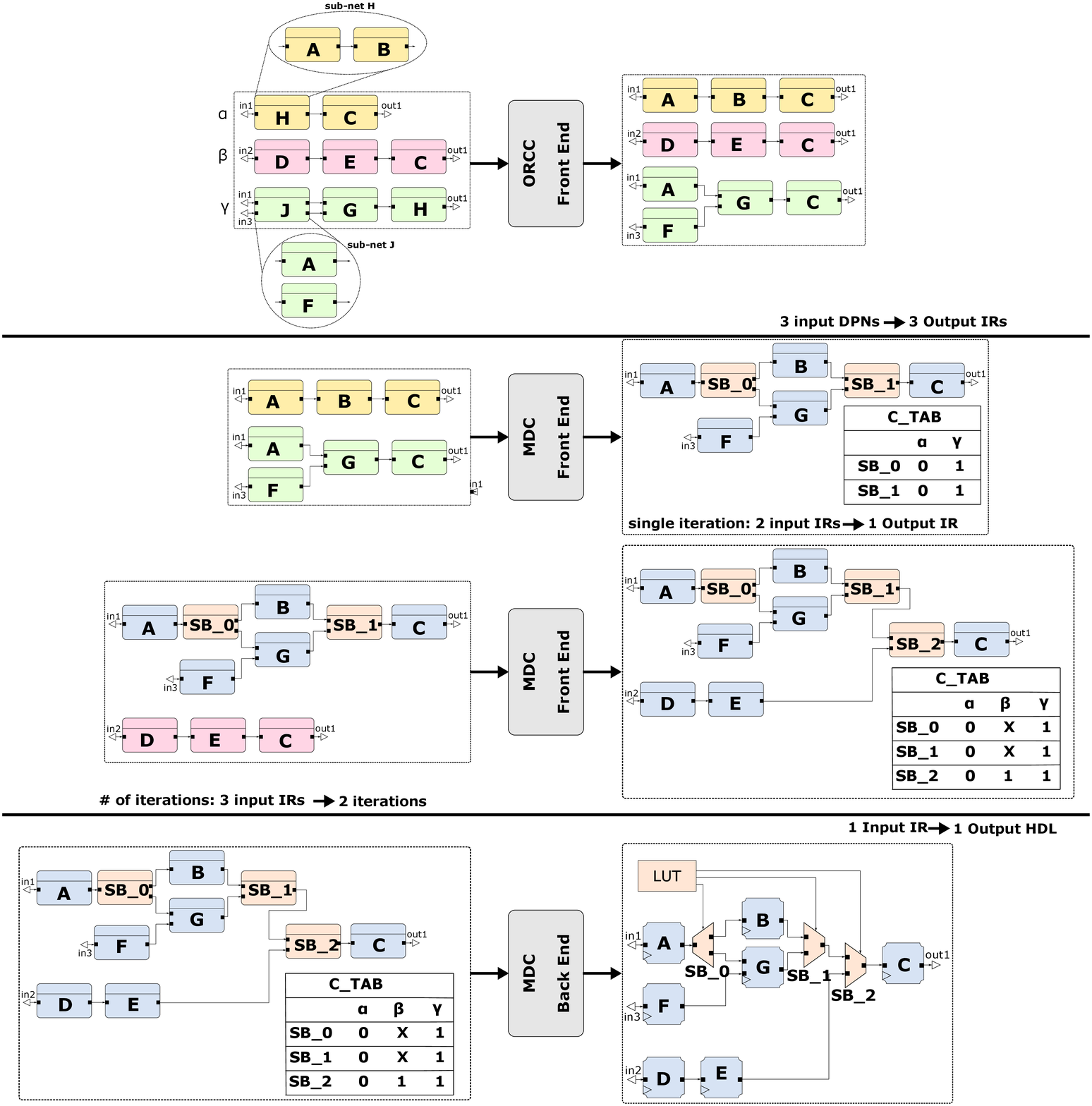}% picture filename
	\caption{Baseline MDC Core: a step-by-step example.}
	\label{fig:mEx}
\end{figure*}

\subsubsection{Step-by-Step Example}
In order to clarify the baseline MDC core functionality Figure~\ref{fig:mEx} \textcolor{black}{illustrates, through a step-by-step example, the iterative datapath merging process of the heuristic algorithm, that derives the reconfigurable multi-functional architecture.}
It considers an example with three different DPN specifications (\emph{$\alpha$},
\emph{$\beta$} and \emph{$\gamma$}), and the generated output is the HDL description of the CGR
architecture. At first ORCC parses the input DPNs, flattens the hierarchical actors and builds the 
corresponding IRs. In particular \emph{$\beta$} is already flattened, being composed
of atomic actors only, while the actor H of \emph{$\alpha$} and the actor J of \emph{$\gamma$}
enclose a sub-network each, and are flattened before proceeding with the merging process.
%%%%%

After parsing  the input DPNs, MDC starts the iterative merging process. MDC front-end
analyses the IRs in pairs to determine which actors can be shared between the two considered networks.
Identical actors are shared in the output IR by introducing dedicated switching elements, 
used to fork ($Sbox\_1\times2$) or re-join ($Sbox\_2\times1$) the path of data. It is important to
notice that for N input IRs, N-1 iterations are required to complete the merging process and, in 
the worst case scenario, the process can end up with N-1 cascaded SBoxes to access a PE shared 
by all the N input DPNs.
%%%%
In the considered case with only three input networks, two iterations are required. In the first run, the merging algorithm identifies actors A and C as identical among \emph{$\alpha$} and 
\emph{$\gamma$}, so it inserts two SBoxes. Then, in the second run, the algorithms identifies
actor C as identical among the previous generated \emph{multi-flow IR} and \emph{$\beta$};
thus, only another SBox is inserted. During each iteration MDC assigns an identification
value to each network and, for each of them, keeps trace of the right selector values to
be assigned to each \emph{SBox}, updating the $C\_TAB$.  

At last the MDC back-end generates the 
\emph{CGR HDL}, mapping the different actors of the \emph{multi-flow IR} 
over the PEs provided within the \emph{HDL components library}. The control signals of the physical 
SBoxes are generated by the LUT module, whose content depends on the final $C\_TAB$ produced by the MDC 
front-end, that guarantees the computing correctness of each input functionality.

\subsection{Structural Profiler}
\label{ss:profilerMDC}

In the adopted iterative merging algorithm MDC processes two networks at a time. 
Since SBoxes are combinatorial elements, a long chain of SBoxes could imply a 
change into the critical path, that may negatively affect the operating frequency. Furthermore, an 
excessive number of switching elements may overcome the benefits of sharing an actor, causing both 
area and static power increment. Therefore, in some cases it would be more efficient to merge only
a subset of the input DPNs. For these reasons, it is fundamental to determine the (sub-)optimal 
design specification(s) that have to be merged into the CGR architectures.

The MDC \textit{Structural Profiler} analyzes all the possible merging configurations, returning the best ones in terms of area, power consumption  and operative frequency~\cite{Palumbo_2016}. %\cite{Palumbo_2013}. 
For each of the different possible DPN merging sequences, the MDC tool extracts the multi-dataflow DFG as described in Section~\ref{s:baselineMDC}. Then, for each possible merging configuration, the MDC \emph{Structural Profiler} computes an implementation cost, based on a back-annotation of the HDL components library coming from area and power consumption estimations of each input DFG. Therefore, given as $M$ the size of the $V$ set of vertices of the graph (the actors), area and power consumption are determined as:

\begin{equation}\label{eq:2}
Area(DFG)=\sum\limits_{i=1}^{M} a_i
\end{equation}
\begin{equation}\label{eq:3}
Power(DFG)=\sum\limits_{i=1}^{M} p_i
\end{equation}

\noindent where $a_i$ and $p_i$ are respectively the estimated area and power of the $i-th$ vertex (actor).
Operating frequency is instead estimated as follows. The \emph{Structural Profiler} $\forall n_i \in InN$ (being $InN$ the set of input DPNs) retrieves the corresponding back-annotated Critical Path (CP, that is the maximum combinatorial delay, responsible of the maximum achievable operating frequency of the circuit), $CP_i$, and defines $CPstatic=max(CP_i)$ as the CP of the non
reconfigurable system configuration (with all the given DPNs in parallel). Then it identifies the longest cascade of SBoxes ($seqSB$) within the considered multi-dataflow $DFG$, that is a combinatorial path since SBoxes are purely combinatorial. 
Given the number of SBoxes ($N_S$) that compose the cascade $seqSB$, and given the number of bits of SBoxes data ($b$), the CP is given by the empirical Equation \ref{eq:4}, where coefficients f(b) and g(b) are technology dependent.

\begin{equation}\label{eq:4}
CP\_seqSB=f(b)*ln(N_S)+g(b)
\end{equation}

\textcolor{black}{The CP of the multi-dataflow DFG, responsible of its maximum achievable frequency ($Freq(DFG)$), is then calculated as the maximum value between the original CP ($CPstatic$) and the CP due to the merging process ($CP\_seqSB$). According to the obtained $Area(DFG)$, $Power(DFG)$ and $Freq(DFG)$, the different possible merging configurations are ranked and optimal solutions, under the different considered metrics, are identified. As already mentioned before, the structural profiler feature is currently available only when ASIC target technology is considered.} For a deeper description of MDC \textit{Structural Profiler}, together with a step-by-step example, please refer to Palumbo et al. \cite{Palumbo_2016}.

\subsection{Dynamic Power Manager}
\label{ss:dynPowerMDC}
In a CGR system all of the logic necessary to compute the 
different functionalities is instantiated in the substrate and the configurations are enabled by 
multiplexing resources in time. When a specific functionality is executed, the rest of the design, that is not involved 
in the computation, is in an idle state.  
As seen in Section \ref{Soa:power}, several techniques can be applied to reduce power consumption and, in some cases, they are automatically implemented by commercial synthesis/place-and-route tools. However, most of the available strategies still require designers to identify the logic to be switched-off and, in some cases, also to specify the power intent files (either UPF or CPF). 

Given the fact that unused resources, in an MDC compliant CGR architecture, can be determined at design-time for any given configuration, it is possible to divide all the resources into sets of  disjointed Logic Regions (\emph{LRs}), composed of resources that are always 
active/inactive together, and reduce their power consumption by applying power saving techniques.

MDC exploits the intrinsic modularity of the dataflow models to automatically identify the minimum set of
\emph{LRs} by applying an identification algorithm that acts at the specification
level (See Algorithm 1 in Palumbo et al. \cite{Palumbo_2016}). The given dataflows are 
analyzed to identify and group together the actors active/inactive at the same time within homogeneous
logic sets. On the MDC Graphical User Interface (GUI) users can choose to enable or not a power-saving strategy.

\subsubsection{Clock Gating}
\label{ss:cg}
MDC exploits the identified \emph{LRs} to implement the clock gating technique that reduces the dynamic power: when a \emph{LR} is not working, its clock
can be turned off to limit the switching activity of the design and, in turn, its power dissipation \cite{Palumbo_2016}.
MDC is able to automatically implement clock gating for either ASIC or FPGA targets. When ASIC target is selected, MDC provides \emph{AND} gate cells that are applied directly on the clock to disable it. Otherwise, if FPGA is selected, MDC instantiates, for each \emph{LR} to be gated, a \emph{BUFG} cell. In the second case, MDC guarantees compatibility with Xilinx design environment and boards only. %(this can be applied only on Xilinx boards).
Targeting FPGAs, the number of \emph{BUFG} cells available on the board is limited. If the
number of identified \emph{LRs} exceeds the amount of available \emph{BUFG} cells, MDC adopts an 
algorithm (see Algorithm 2 in \cite{Palumbo_2016}) to reduce the number of gateable \emph{LRs}.% , identifying the sub-optimal set of \emph{LRs}, where switching activity in unused FUs is still present. 

\subsubsection{Power gating}

The clock gating acts on the dynamic power. However, as transistors get smaller, it is no longer possible to neglect the contribution of the static power. One of the most popular techniques to reduce the consumption of static power is the power gating. The main idea behind it is the same as the one of clock gating: when a portion of the design is not involved in the computation, it can be switched-off, by means of a sleep transistor. As the clock gating, the power gating can be applied to the \textit{LRs} identified by MDC, as demonstrated in \cite{Fanni_2015}. However, clock gating can be handled almost easily during the design and implementation process; while power gating is a more invasive technique, since it requires the insertion of several extra logic to handle the inter-block communication and the powering down/up transitions.

Firstly, it is required the insertion of the \emph{sleep transistors} (or \emph{power switches}) between the gated region (or \emph{power domain}) and the main power supply to selectively switch on/off the power supply of the region. However, this is not enough to handle the correct power-down/up sequence, which includes also the isolation of signals from the shut-down domain. 
The power domains to be powered-down have to be isolated before power is switched off, and have to remain isolated until the power is again totally on. The \emph{isolation logic} is typically used between the powered-down region and the powered-on ones, to avoid the transmission of spurious signals in input to powered-on cells.
In certain cases, the state of registers needs to be maintained to guarantee the proper operation of the system, when the regions are powered-on. For this purpose, \emph{state retention logic} has to be adopted. Retention cells typically have a low power consumption shadow register, connected to the main power supply, where the state of the main register is saved when the corresponding region is powered-down. 

All this additional logic can be manually inserted by the designers in the RTL architecture or through a \textit{power format file}. Manual definition is highly error prone: it requires modelling the impact of power during simulation and providing multiple definitions for synthesis, placement, verification and equivalence checking~\cite{lowPowGuide}. A power format file allows designers to specify the power intent early in the design and without any direct modification of the RTL code. The two most commonly used low power flows are the UPF~\cite{IEEE:UPF} and the CPF~\cite{SI2CPFspecification}.
In MDC, the CPF is adopted. %The CPF can be divided into two main parts: technology and power intent. 
%\begin{itemize}
%	\item The technology part sets the timing libraries and specifies the low-power cells to be used within the technology specific physical libraries.
%	\item The power intent part depends on the design and manages the power domains, the power modes and the respective transitions. Here the power cells, specified on the technology part, are instantiated and associated to the logic in the design, accordingly with the power domains to which they belong. 
%\end{itemize}
%The capability of MDC of applying automatically the clock gating methodology (which acts only on dynamic power) has been extended to implement also a coarse-grain power gating strategy for ASIC designs \cite{Fanni_2015}. 
To apply the power gating, firstly the MDC \textit{Logic Regions Identification} algorithm (see Algorithm 1 in \cite{Palumbo_2016}) has been modified to include also the switching modules (SBoxes) of the CGR system that, being combinatorial, were not included in the regions to be clocked-off (see Algorithm 3 in \cite{Palumbo_2016}). At the end of the process, each \emph{LR} is mapped into a different power domain. This implies creating, for each \emph{LR}, a power domain into the CPF file, defining for each switchable domain also the shut-off condition. Instances belonging to each domain are related to the actors that belong to the corresponding \emph{LR}. Then, to give the information about the power specification to the synthesizer, MDC has been extended to automatically generate also the CPF file. For a deeper explanation of the automatic application of power gating in MDC, and for step-by-step examples, please refer to Palumbo et al. \cite{Palumbo_2016}.

\subsubsection{Hybrid Clock/Power Gating}
The power gating is a technique that can be extremely beneficial in saving both static and dynamic power. However, as described above, it is a quite invasive technique that requires several additional logic, and blindly shutting-off the idle logic is not always the best strategy. In some cases the power consumption due to the power saving logic might exceed the amount of power saved by switching-off the idle logic, i.e. in small idle regions. In other cases, power gating could turn out to be less effective than clock gating, i.e. in those regions where sequential logic is predominant. For these reasons, it could be useful to determine, at an early design stage, which regions may benefit from power saving application and, also, to correctly identify in advance which techniques should be used in each of them individually.
%analysing the design to identify which portions of it can benefit from power saving application, also identifying the best technique, and which not.

%When the architecture is manually designed and implemented and only few idle areas are involved, such an analysis is affordable. However, talking about CGR systems, there could be tens of regions that are idle during the execution of one or more configurations. In such a case, identifying the best power saving strategy could be highly time consuming. 
To overcome the limits of a blindly applied unique power management strategy, in MDC it has been adopted a power estimation flow capable of characterizing, at a high-level of abstraction, the \emph{LRs} identified by the MDC power extension, and to estimate power and clock gating overhead before any physical implementation. The estimation is based on two sets of models that determine the static and dynamic consumption of each \emph{LR} when clock gating or power gating are applied. The proposed models are derived after a single logic synthesis of the baseline CGR system generated by MDC, carried out with commercial synthesis tools from the analysis of the power reports obtained after netlist simulation \cite{Fanni_2016, Palumbo_2015}. 

\subsection{Coprocessor Generator}
\label{ss:coprMDC}

\begin{figure*}[t!h]
	\centering
	\includegraphics[width=0.7\textwidth]{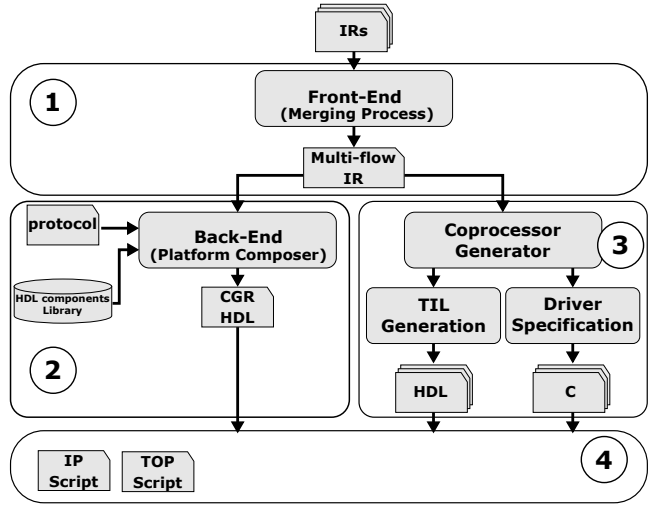}
	\caption{Design flow overview.}
	\label{fig:copr_flow}
\end{figure*}

MDC tool was already able of automatically composing, synthesizing and deploying runtime reconfigurable coprocessors. In its first version, generated coprocessors were compliant with Xilinx ISE Design Suite \cite{Sau_2015}. In this paper it is presented the new MDC \emph{Coprocessor Generator} flow compliant with the Xilinx Vivado design suite. \textcolor{black}{A detailed discussion of the main differences and improvements introduced in this work with respect to the previous version of the coprocessor generator is provided in Section~\ref{ssect:diff}.} Figure~\ref{fig:copr_flow} illustrates the new MDC \emph{Coprocessor Generator} flow. MDC generates the multi-dataflow (\emph{Multi-flow IR}) merging the input dataflow specifications as described in Section~\ref{s:baselineMDC} (1). Then, starting from the generated multi-dataflow network, MDC composes the corresponding CGR core (2). In parallel, it generates also the files and the necessary logic to embed the computing core into a configurable Template Interface Layer (TIL) (3). Finally, to easily deploy and use the coprocessor, MDC provides the Xilinx Vivado scripts to automatically pack the logic into a processor-coprocessor architecture and the software drivers to ease its use (4). 

%Since the coprocessor is strongly dependent on the context and on application model, that is the input dataflow specifications, 
Several options are available to the user in order to maximize efficiency of the obtained result. In particular, it is possible to choose:
\begin{itemize}
\item the kind of host processor;
\item the processor-coprocessor coupling;
\item the adoption of DMA engines.
\end{itemize}
Each of these aspects impacts on different steps of the coprocessor generation flow: the TIL generation is affected only by the coupling, while processor, coupling and DMA preferences directly impact on drivers and scripts generation. In the following we are going to describe more in detail such steps and their dependency on the user choices. %in terms of processor, coupling and DMA.

\subsubsection{Template Interface Layer}
\label{ssect:til}
% inizio paragrafo copiato da reconfig (leggermente modificato)
Generally speaking, coprocessing units can have different degrees of coupling with the host processor. A loosely coupled coprocessor is far from the processor, it is typically accessible through the system bus and it is affected by medium/high communication latency for both control and data transfers. A tightly coupled coprocessor is close to the processor and it often shares with the processor high-level memories. A loosely coupled coprocessor can be easily adopted in different contexts, since it is connected to a generic system bus. On the contrary, it is hard to extend the adoption of a tightly coupled coprocessor to different systems, since it has dedicated links and memory accesses.
% fine paragrafo copiato da reconfig (leggermente modificato)
MDC supports two different levels of coupling that exploit the AMBA AXI4 communication protocol \cite{Xilinx:axi}. Users can choose between:
\begin{itemize}
	\item memory-mapped TIL (mm-TIL): a memory-mapped loosely coupled coprocessor; 
	\item stream-based TIL (s-TIL): a stream-based tightly coupled coprocessor. 
\end{itemize}

\begin{figure}[!t]
	\centering
	\includegraphics[width=0.8\textwidth]{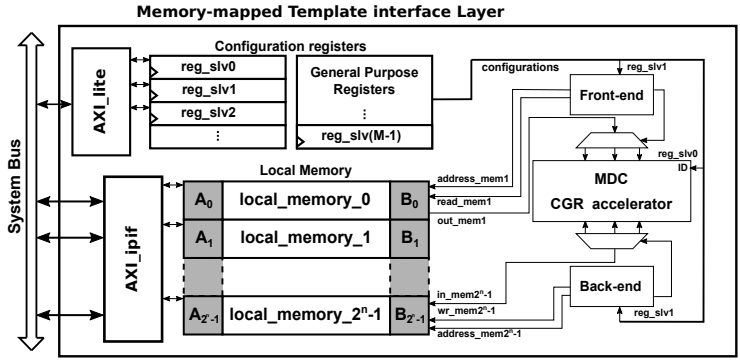}
	\caption{Architecture of the memory-mapped Template Interface Layer (mm-TIL).}
	\label{fig:mm_TIL}
\end{figure}

Figure~\ref{fig:mm_TIL} shows the architecture of the mm-TIL whose main blocks are: the configuration registers bank, a local memory and a front-end or a back-end for each I/O port. 
%%%%%%%%% copied by carlo's thesis
The local memory contains all the data to be processed by the coprocessor and the computed results.  It has to be fully written by the processor before the coprocessor execution phase and it has to be fully read once the coprocessor has completed the task. A dedicated address range of the processor is reserved to the local memory.
%%%%%%%%%%
The memory banks are written through the AXI4-full (AXI\_ipif in Figure~\ref{fig:mm_TIL}), generally used for high performance memory-mapped requirements.
The configuration registers bank is the entity in charge of storing the configuration of the coprocessor. The configuration includes the ID of the kernel (corresponding to the input dataflow) to be executed and the data number for each I/O port. The data number is the amount of data to be read/written from/to the local memory. The configuration registers are written through the AXI4-Lite (AXI\_lite in Figure~\ref{fig:mm_TIL}) interface that is generally used for simple, low-throughput memory-mapped communication. The front-end is responsible for the data transfer from the local memory to the reconfigurable computing core, while the back-end transfers the processed data from the reconfigurable computing core to the local memory.

\begin{figure}[!t]
	\centering
	\includegraphics[width=0.6\textwidth]{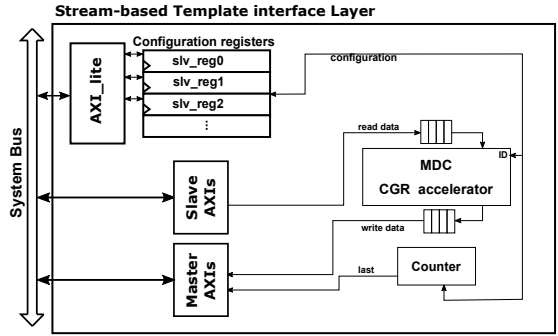}
	\caption{Architecture of the stream-based Template Interface Layer (s-TIL).}
	\label{fig:s_TIL}
\end{figure}

Figure~\ref{fig:s_TIL} depicts the s-TIL architecture that leverages on the AXI4-Stream communication protocol, generally used for high speed streaming data transfers. The configuration registers bank, as in the mm-TIL, saves the coprocessor configuration. In the s-TIL the front-end and back-end are not present since the AXI4-Stream interfaces are directly connected to the reconfigurable computing core I/O ports. However, in order to properly derive the AXI4-Stream \texttt{last} signal, it is necessary to insert a counter for each output port. 

\subsubsection{Driver Specification}

At an higher-level of abstraction, the software drivers offer an interface that masks the system configuration complexity, providing a C function for each configuration of the CGR coprocessor. Taking care of the processor-coprocessor communication, such step of the coprocessor generation is affected by the user choices in terms of processor, coupling and DMA. In particular, the coupling and DMA change the way data is transferred from/to the processor to/from the coprocessor. The processor choice also influences such transfers, since the two admitted possibilities are strongly different:
\begin{enumerate}
\item MicroBlaze is a soft-core instantiated in the programmable logic that offers strong customization (it can support direct stream communication) at the price of performance (it is slower, and it has smaller memories with fast access);
\item ARM is a hard-core present only in some target devices (Zynq-7 family), capable of delivering strong performance (it is faster, and it has big memories with quick access), but limited customization (it does not support direct stream communication).
\end{enumerate}

Listing~\ref{list:int_dr} in~\ref{app:list} shows the prototype of the driver top functions for both memory-mapped and stream-based coprocessors, and for one possible configuration of the CGR substrate. The driver top functions have two arguments per reconfigurable computing core I/O port, \texttt{data\_<port\_name>} and \texttt{size\_<port\_name>}, that are respectively data pointer to load (or store) data to (or from) an input (or output) port, and the number of data related to that port. In the considered example there are three ports: \texttt{in\_size}, \texttt{in\_pel} and \texttt{out\_pel}. It is clear as the interfaces for the two cases, memory-mapped and stream, are identical. This allows software designer with little knowledge of hardware design to easily use the generated processor-coprocessor systems, without considering the underlying processor-coprocessor coupling. 
Then the body of the function manages communication between the host processor and the coprocessor (see Listing~\ref{list:bodyDr} in~\ref{app:list}). 
For each I/O port of the reconfigurable computing core, a configuration word is written into the proper configuration register in order to make the coprocessor aware of the amount of data expected for such port (\texttt{*(config + 1) = size\_<port\_name>}). Please note that in stream-based coprocessor this is not necessary for input ports. Then, the indicated amount of data (\texttt{size\_<port\_name>}) for each input port involved in the current computation is sent to the corresponding local memory or to the input FIFO according to the chosen coupling, memory-mapped or stream-based (see lines under \texttt{//send data port in\_size} comment) and to the fact that DMA engines have to be adopted or not. At last, the processor can read back the results into the processor from the output ports (see lines under \texttt{//receive data port out\_pel} comment). In the case of memory-mapped coupling, the processor needs to monitor through polling a configuration register where a \texttt{done} flag is stored at the end of the computation. In the case of stream coupling a \texttt{done} flag is not necessary, since the processor only needs to evaluate the state of the output FIFOs.

\subsubsection{Coprocessor Deployment}

 In order to integrate and deploy the peripheral as a standard Xilinx IP, MDC provides an automatic script for Xilinx Vivado design suite (see Listing \ref{list:ip_scr} in~\ref{app:list}). The inputs for the script are the HDL description of the generated TIL, including TIL submodules (config registers, local memories, front end, \dots) and the CGR core modules (\texttt{add\_files \$hdl\_files\_path}), any required HDL library (\texttt{set\_property library caph}) and the generated drivers (\texttt{ipx::add\_file\_group -type software\_driver}). The output is the resulting Xilinx ready-to-use IP comprehensive of software drivers. This means that, once added to the IP catalog of a certain project, it could be added and manipulated by all the Vivado common features adopted to develop heterogeneous systems, such as block design and Software Development Kit export.

MDC also provides another script to instantiate the generated IP into an integrated processor-coprocessor system, within the Vivado environment (see Listing  \ref{list:top_scr} in~\ref{app:list}). According to the user choice, the host processor can be a hard-core (ARM processor) or a soft-core (Microblaze); in the considered example an ARM processor is instantiated (\texttt{create\_bd\_cell -type ip -vlnv ... processing\_system7\_0}). The communication between processor and coprocessor can be managed either with or without DMA modules. In general, the kind and number of DMA modules adopted in the specific integrated system depend on the processor and on the coupling between processor and coprocessor. In the considered example, for a memory-mapped communication between an ARM core and the coprocessor through DMA, the AXI Central DMA (AXI CDMA) module is instantiated (\texttt{create\_bd\_cell -type ip -vlnv ... axi\_cdma\_0}). 

The user choices %, the kind of host processor (ARM or MicroBlaze), the DMA usage (yes or not) and the processor-coprocessor coupling (memory-mapped or stream), 
strongly influence all the integrated system scripts and the resulting system:
\begin{itemize}
    \item The kind of processor indeed is playing a role on the logic necessary to manage stream communication since, in the ARM case, it is not supported directly and requires additional modules to allow memory-mapped to stream conversion (AXI-Stream FIFOs or AXI AXI DMA modules). Besides that, processors impact on the system performance, more precisely in terms of software execution speed and memory availability.
    \item The DMA usage allows more efficient data transfers, while introducing an overhead in terms of resources. Its adoption should then be limited to those cases where lots of data have to be transferred from/to the host processor to/from the coprocessor.
    \item The level of coupling between processor and coprocessor plays a role in the resource versus performance trade-off, as explained more in detail in Section~\ref{ssect:til}. This has also effects on the glue logic necessary to let the two interlocutors talk together, that is bus systems (AXI Interconnect), FIFOs and DMAs, when used.
\end{itemize}
According to the described degrees of freedom, the integrated processor-coprocessor system can involve several other Xilinx IPs. Table~\ref{tab:xil_ips} summarizes the kind and number of additional Xilinx IPs required for each possible scenario. Please note that the number of such additional IPs is sometimes depending on the number of I/O ports of the reconfigurable computing core (e.g. the FIFOs for the stream-based coupling possibilities). Thus, the amount of resources can easily grow if the dataflow models of the applications to be accelerated have lots of I/O. In general, the choice in terms of processor, coupling and DMA is depending on performance and resource requirements, but also on the starting dataflow models, as we are going to demonstrate in Section~\ref{s:ass}.

\begin{table}
 \begin{center}
 \caption{Additional Xilinx IPs, besides the processor and the coprocessor, required for each possible scenario of the MDC coprocessor generator. %\textcolor{blue}{maybe insert refs to Xilinx doc?}
 }
 \begin{tabular}{ccc||l}
 \textbf{processor} & \textbf{coupling} & \textbf{DMA} & \textbf{additional Xilinx IPs} \\
   \hline
   \hline
  MicroBlaze & mm & no & AXI4 Interconnect \\
  \hline
  \multirow{2}{*}{MicroBlaze} & \multirow{2}{*}{mm} & \multirow{2}{*}{yes} & AXI4 Interconnect \\
  & & & AXI DMA\\
  \hline
  \multirow{2}{*}{MicroBlaze} & \multirow{2}{*}{stream} & \multirow{2}{*}{no} & AXI4 Interconnect \\
  & & & AXI4-Stream Data FIFO (1 per I/O port) \\
  \hline
  \multirow{3}{*}{MicroBlaze} & \multirow{3}{*}{stream} & \multirow{3}{*}{yes} & AXI4 Interconnect \\
  & & & AXI4-Stream Data FIFO (1 per I/O port) \\
  & & & AXI CDMA (1 per I/O port) \\
   \hline
   ARM & mm & no & AXI4 Interconnect \\
   \hline
   \multirow{2}{*}{ARM} & \multirow{2}{*}{mm} & \multirow{2}{*}{yes} & AXI4 Interconnect \\
   & & & AXI DMA \\
   \hline
   \multirow{2}{*}{ARM} & \multirow{2}{*}{stream} & \multirow{2}{*}{no} & AXI4 Interconnect \\ 
   & & & AXI-Stream FIFO (1 per couple of I/O ports) \\
   \hline
   \multirow{3}{*}{ARM} & \multirow{3}{*}{stream} & \multirow{3}{*}{yes} & AXI4 Interconnect \\
   & & & AXI4-Stream Data FIFO (1 per I/O port) \\
   & & & AXI CDMA (1 per couple of I/O ports) \\
 \hline
 \hline
 \end{tabular}
 \label{tab:xil_ips}
 \end{center}
 \end{table}

\subsubsection{Main Improvements with Respect to \cite{Sau_2015}}
\label{ssect:diff}
\textcolor{black}{MDC Coprocessor Generator was firstly introduced in~\cite{Sau_2015}. Substantial improvements have been introduced in the current work, resulting from an almost complete re-engineering of such advanced feature of MDC. The improvements involve several aspects of the Coprocessor Generator, from technical to compatibility ones. Here, a detailed list of them is provided:
\begin{enumerate}
    \item the targeted Xilinx design environment has been updated from ISE to Vivado, leading to big advantages in terms of system integration, being Vivado the standard de facto for users adopting devices of this vendor;
    \item the supported host cores are now two: the Xilinx MicroBlaze soft-core, already supported in~\cite{Sau_2015}, and the ARM hard-core, one of the leading embedded core architectures worldwide;
    \item the supported system buses have changed from Xilinx proprietary Processor Local Bus (PLB), for memory-mapped coupling, and Fast Symplex Link (FSL), for stream coupling, to ARM AMBA AXI4 system bus. These latter are adopted in target devices of different vendors and providing several protocols. MDC coprocessors currently supports small register-oriented memory-mapped transfers (AXI4-Lite), big memory-mapped transfers (AXI4-Full), and stream transfers (AXI4-Stream);
    %the supported system buses have changed from Xilinx proprietary Processor Local Bus (PLB), for memory-mapped coupling, and Fast Symplex Link (FSL), for stream coupling, to ARM AMBA AXI4 system bus, adopted in target devices of dierent vendors and providing several protocols to small register-oriented memory-mapped transfers (AXI4-Lite), big memory-mapped transfers (AXI4-Full) and stream transfers (AXI4-Stream);
    \item the accelerator interfaces have been made more resource efficient since now configuration and parameters are sent through a reduced memory-mapped interface (AXI4-Lite) instead of adopting a standard data transfer (memory-mapped or stream) interface, as occurred in~\cite{Sau_2015};
    \item the reconfigurable computing core can now adopt a generic hardware communication protocol (see Section~\ref{ss:caph}), specified through a dedicated input file, and the glue logic to communicate with the system bus (a finite state machine in the memory-mapped case, simple logic gates in the stream one) is shaped accordingly (previously, only the RVC-CAL hardware communication protocol was supported);
    \item the adoption of DMA engines has now been integrated in the automated flow. % meaning that it can be now included in the overall processor-coprocessor system and the drivers can be generated to configure DMA accordingly. 
    In the previous version, it was possible to integrate the generated coprocessors in systems with DMAs, but designers were requested to manually make the system integration and to provide driver modification to configure DMAs;
    \item the system integration has been now automated through two TCL scripts, one for packing the coprocessor as a standard Xilinx Vivado IP and the other for building the processor-coprocessor system, which are directly processable by the targeted Xilinx Vivado Environment (scripts for system integration automation were not provided at all in~\cite{Sau_2015}, resulting in an additional effort required to the user especially when coprocessors had lots of bus interfaces);
    \item fostering interoperability among MDC and other complementary tools (see Section~\ref{sect:interf} for more details), the MDC system deployment capabilities have been extended to: %Coprocessor Generator has been made able to:
    \begin{itemize}
        \item differentiate % differentiating
        inputs of the reconfigurable computing core between standard dataflow inputs, linked with massive data transfer interfaces (AXI-Full or AXI-Stream), and dynamic parameters, linked with the lightweight AXI-Lite interface to serve as knobs for the SPIDER run-time management (see Section~\ref{ss:spider});
        \item instantiate  %instantiating
        Performance Monitoring Counters (PMCs) to keep trace of interesting events during execution, as well as generating an XML file with the PMCs info to be passed as input to PAPIFY, a tool taking care about PMC triggering and data gathering at run-time leveraging on a standard generic HW/SW interface (see Section~\ref{ss:papi});
        \item generate a CGR substrate, compliant with ARTICo\textsuperscript{3} acceleration slots, thus delivering a multigrain reconfigurable platform by combining the delivered CGR with the DPR provided by the same ARTICo\textsuperscript{3} (see Section~\ref{ss:artico}). %\textcolor{blue}{Tizi controlla/aggiusta frase se non va bene TCL}. \textcolor{blue}{Tiziana: Ragazzi, questo non era corretto, quello script non esiste più, ora c'è un backend apposito per artico, che gira a prescindere dal coprocessor generator (di fatto per come è costruito mdc non è nemmeno possibile far andare contemporaneamente i due). La classe del backend è un'estensione della classe del coprocessor generator, l'ho cambiato e di conseguenza ho cambiato un filino la parte che introduce questo elenco puntato. rileggete.}
    \end{itemize}
\end{enumerate}
Please note that most of these improvements, such as compatibility/interoperability ones (1, 2, 3, 5, 8) are not measurable, but they favour the adoption of the tool by a wider public or for more complex and complete purposes. Other improvements (6, 7, 8), dealing with flow automation, are difficult to be measured as well, being strongly user-dependent. If the second term of comparison is an expert hardware designer or a software developer, the evaluated metric would be completely different. Nevertheless, the benefits of automation are generally irrefutable. Technical improvements (4, 8) could be measurable, but some results may be trivial. Adopting AXI-Full or AXI-Stream interfaces, which are conceived for data intensive transfers, for performing few single data transfers is by definition worse than opting for minimal AXI-Lite ones. Please note that, this last improvement on interfaces is overcoming a known issue already pointed out in \cite{Sau_2015}.
}

\section{Assessment}\label{s:ass}
This section provides an assessment of the latest features introduced in the MDC tool considering a robotic test case, belonging to a completely new application scenario for MDC. To assess such features, the customization possibilities for CGR accelerators will be shown and analyzed under several aspects.

\subsection{Reference Application and Designs Under Tests}
In this paper, to demonstrate the usage and potentials of the MDC tool, the \emph{Damped Least Square} (DLS) algorithm \cite{Buss2004, Buss2009} is adopted. The DLS solves \emph{Inverse Kinematics} (IK) problems and, in the present case, is used to implement the controller of a robotic arm implemented over an FPGA device. Given the assumption that any robotic manipulator is composed of different parts, namely: \emph{(i)} the base, \emph{(ii)} the rigid links, \emph{(iii)} the joints (each of them connecting two adjacent links), and \emph{(iv)} the end effector, to control its movement, it is possible to compute the joint angles that will bring the end-effector in the desired position. 

\begin{figure}[t!h]
    \centering
    \includegraphics[width=\textwidth]{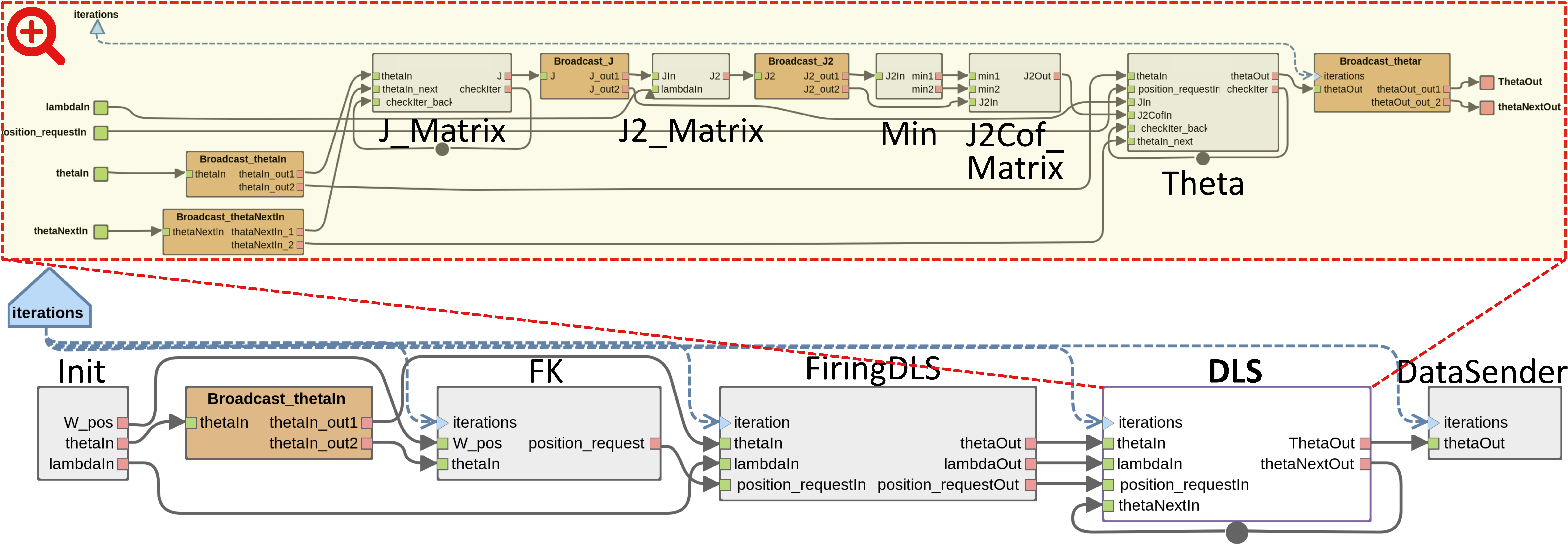}
    \caption{Hierarchical PiSDF description of the DLS in PREESM. \textcolor{black}{Grey and orange boxes correspond to the actors and routing blocks (broadcasts) respectively, while the blue pentagon represents the PiSDF parameter (named \textit{iterations}).}}
    \label{fig:pisdf}
\end{figure}

The dataflow specification of the DLS we are using as a starting point for our assessment is depicted in Figure~\ref{fig:pisdf}. DLS is an iterative algorithm that splits a given trajectory in different sub-sections, each one calculated separately and in a sequential way. The number of iterations, that is the number of sub-sections of the trajectory, is chosen according to the tolerance, defined as the error of the obtained end-effector position with respect to the desired one, and to the trajectory length. \textcolor{black}{This parameter determines how many times the DLS block in Figure \ref{fig:pisdf} is executed. In particular, all the tasks belonging to the DLS algorithm are performed \textit{iterations} times. Among them, the blocks \textit{J\_Matrix}, \textit{J2\_Matrix}, \textit{Min}, \textit{JCof\_Matrix} enable to obtain a matrix derived from the Jacobian matrix J, as explained in \cite{Buss2004}, which is needed to compute the joint space vector in the last task \textit{Theta}. %On the other hand, looking at same hierarchical level of the DLS subgraph, 
The other blocks enable handling input and output data. \textit{Init} provides the starting angles of the joints, the desired final point and the $\lambda$-factor, which is used in literature to manage singularities in the workspace. \textit{FK} evaluates the end-effector position by using the Forward Kinematics. \textit{FiringDLS} handles DLS executions management. \textit{DataSender} transmits all the computed angles to the robotic arm. A more detailed description of the DLS algorithm implementation and its blocks goes beyond the scope of this paper and can be found in a previous work by Fanni et al. \cite{FanniSRSTP19}.} 

Starting from the original MATLAB description of the DLS algorithm, we modelled the above presented PiSDF description in PREESM. Since the \emph{Actors} of the PiSDF graph are written in C language, their hardware counterparts (described as HDL codes) have been obtained by synthesizing them through Vivado HLS.
%we modelled a high-level dataflow description compliant with the MDC tool. 
From this dataflow we derived two different configurations of the application, as defined hereafter:

\begin{itemize}
    \item \emph{TOP\_BL} is the baseline (BL) version of the DLS algorithm, implemented with the following set of actors: \emph{J\_Matrix}, \emph{J2\_Matrix}, \emph{Min}, \emph{J2Cof\_Matrix} and \emph{Theta};
    \item \emph{TOP\_HP} is the high performance (HP) version of the DLS algorithm, implemented with the following set of actors: \emph{J\_Matrix\_HP}, \emph{J2\_Matrix\_HP}, \emph{Min}, \emph{J2Cof\_Matrix} and \emph{Theta}.
\end{itemize}

A reconfigurable datapath (\emph{Reconf}) has been automatically generated by applying the MDC baseline merging flow to \emph{TOP\_BL} and \emph{TOP\_HP}. \textcolor{black}{In parallel, standalone implementations of the \emph{TOP\_BL} (\emph{Stand\_BL}) and \emph{TOP\_HP} (\emph{Stand\_HP}) configurations have been developed to provide a term of comparison for the reconfigurable datapath.}

\subsection{Datapath Level Results}
In this section results related to the above presented datapaths are shown.
Table~\ref{tab:stdln_res} reports on resource occupancy at actors level and at system level. Actors level results come from Vivado HLS reports; while, system level ones from Vivado synthesis reports. Data have been retrieved with an operating frequency of 100 MHz and targeting a Zynq-7000 device (XC7Z020CLG484). At system level, \emph{Stand\_BL}, \emph{Stand\_HP} and \emph{Reconf} dataflow implementations are compared. %, this latter built by combining \emph{TOP\_BL} and \emph{TOP\_HP}. 
\emph{Reconf} is capable of executing both profiles of the DLS algorithm, multiplexing actors in time according to the current requirements. As detailed below, the price to pay for being able to implement several profiles, surfing among execution trade-offs on a CGR substrate, is certainly  a higher resource usage. %The goal here was to try to implement several profiles with different execution trade-offs on a CGR substrate. 

As shown in the Table~\ref{tab:stdln_res}, the two execution profiles \emph{BL} and \emph{HP} differ only for \emph{J\_Matrix} and \emph{J2\_Matrix} actors: in the \emph{BL} profile they are present in a baseline version, while in the \emph{HP} one they have been replaced with their high performance version capable of accelerating computation, but employing more resources on the FPGA. Please note that \emph{J\_Matrix} and \emph{J\_Matrix\_HP} correspond to the most computationally intensive actors and are responsible for the main resource demand in both cases. \textcolor{black}{Standalone implementations of the two execution profiles (\emph{Stand\_BL} and \emph{Stand\_HP} rows) give also an idea of their differences in terms of complexity. Of course, being able to execute both profiles, \emph{Reconf} is requiring more resources than the isolated, mono-profile, \emph{Stand\_BL} and \emph{Stand\_HP}. %, which can, on the contrary, execute only one out of two profiles.
Considering an additional solution where the two standalone profiles are put in parallel (\emph{Stand\_BL + Stand\_HP} row) to have a system with the same execution capabilities of \emph{Reconf}, resource usage is higher, motivating than the needs of applying CGR.} It is also possible to notice that the standalone and \emph{Reconf} systems have an overall resource demand, in terms of LUTs and Flip-Flops (FF), that is lower than the sum of their actors, since FPGA slices that are partially occupied by different actors are merged together in the overall systems to maximize efficiency. 

\begin{table}
 \begin{center}
 \caption{Resource occupancy of the generated reconfigurable datapath and of the involved actors.}
 \begin{tabular}{r||cccc|cc}
 \multirow{2}{*}{\textbf{actor/network}} & \multirow{2}{*}{\textbf{LUT}} & \multirow{2}{*}{\textbf{FF}} & \multirow{2}{*}{\textbf{DSP}} & \multirow{2}{*}{\textbf{BRAM}} & \multicolumn{2}{c}{\textbf{profile}}  \\
   &  &  &  &  & \textbf{BL} & \textbf{HP} \\
 \hline
 \hline
 \emph{J\_Matrix}      & 7051  & 3852  & 18 & 15 & x & \\
 \emph{J\_Matrix\_HP}  & 23707 & 12627 & 88 & 17  &   & x \\
 \emph{J2\_Matrix}     & 1697  & 986   & 5  & 0 & x &  \\
 \emph{J2\_Matrix\_HP} & 2621  & 1600   & 11  & 0 &  & x \\
 \emph{Min}            & 1108   & 579   & 5  & 0 & x & x \\
 \emph{J2Cof\_Matrix}  & 2662  & 1628  & 8  & 2 & x & x \\
 \emph{Theta}          & 2041  & 1263  & 5  & 0 & x & x \\ 
 \hline
 \textcolor{black}{\emph{Stand\_BL}} &  8242 & 7164 & 41 & 8 & x &  \\
 \textcolor{black}{\emph{Stand\_HP}} &  17515 & 17176 & 138 & 7 &  & x \\
 \textcolor{black}{\emph{Stand\_BL} + \emph{Stand\_HP}} &  25757 & 24340 & 179 & 15 & x & x \\
 \hline
 \emph{Reconf}        &  22283 & 21057 & 161 & 14 & x & x \\
 \hline
 \hline
 \end{tabular}
 \label{tab:stdln_res}
 \end{center}
 \end{table}

Table~\ref{tab:stdln_toff} depicts the achieved latency versus power trade-offs on the \emph{Reconf} system. Data have been retrieved with Vivado running post-synthesis simulations at 100 MHz. To obtain better power estimations, switching activity gathered during the same post-synthesis simulations has been considered. The table shows how the baseline \emph{BL} profile is slower, being able to complete one iteration of the DLS algorithm in more than 63 us, but consumes a small amount of power, that is around 0.25 W. The high performance \emph{HP} profile is nearly twice faster, concluding the computation of one DLS iteration in about 33 us, while requires a higher amount of power, 0.27 W. Thus, a trade-off between execution latency and consumed power is present in the reconfigurable datapath developed with MDC. \textcolor{black}{Table~\ref{tab:stdln_toff} highlights also the difference between reconfigurable and standalone systems. In terms of execution latency, being implemented with the same actors, \emph{Reconf} in \emph{BL} profile requires the same time as \emph{Stand\_BL} to compute the considered trajectories. The same occurs for \emph{HP} profile. Differently from latency, power consumption varies going from standalone to reconfigurable. In fact, standalone executions of the two profiles consume less than their execution on the reconfigurable system, due to the fact that in standalone systems only the required resources are present. In reconfigurable system, during the execution of one of the two supported profiles, resources involved only in the other profile waste power. This contribution could be reduced by exploiting the MDC Dynamic Power Manager during the \emph{Reconf} system generation, as explained in Section~\ref{ss:dynPowerMDC}. In this case we intentionally did not use such MDC feature to highlight the difference. Please also note that, even if MDC Dynamic Power Manager has not been used, the \emph{Reconf} system executing \emph{BL} profile consumes less than \emph{Stand\_HP} system, strengthening the motivation towards the usage of such a reconfigurable solution.}

\begin{table}[!htb]
    \caption{Trade-off between latency and power for the generated reconfigurable datapath running with the two execution profiles and for the two standalone designs (tested trajectories 1, 2, 3 and 4 require respectively 250, 200, 100 and 200 iterations). \textcolor{black}{Please note that reconfigurable and standalone results in terms of latency are equal for both the considered \emph{BL} and \emph{HP} profiles.}}
    \label{tab:stdln_toff}
    \begin{minipage}{.5\linewidth}
     % \caption{Latency values for each execution profile.}
      \centering
        \begin{tabular}{r||cc}
            \multicolumn{3}{c}{\textbf{Latency [ms]}} \\
            \multirow{2}{*}{\textbf{task}} & \multicolumn{2}{c}{\emph{Reconf/}\textcolor{black}{Stand}} \\
            & \textbf{BL} & \textbf{HP} \\
            \hline
            \hline
            1 iteration & 0.0633 & 0.0334 \\
            \hline
            trajectory 1 & 15.83 & 8.34 \\
            trajectory 2 & 12.67 & 6.67 \\
            trajectory 3 & 6.33 & 3.34 \\
            trajectory 4 & 12.67 & 6.67 \\
            \hline
            \hline
        \end{tabular}
    \end{minipage}%
    \begin{minipage}{.5\linewidth}
      \centering
       %\caption{Power values for each execution profile.}
        \begin{tabular}{r||cc|cc}
            \multicolumn{5}{c}{\textbf{Power [mW]}} \\
            \multirow{3}{*}{\textbf{kind}} & \multicolumn{2}{c|}{\emph{Reconf}} & \multicolumn{2}{c}{\emph{\textcolor{black}{Stand}}} \\
            & \multicolumn{2}{c|}{\textbf{profile}} & & \\
            & \textbf{BL} & \textbf{HP} & \textcolor{black}{\emph{HP}} &  \textcolor{black}{\emph{BL}}\\
            \hline
            \hline
            total   & 246 & 268 & 171 & 251 \\
            \hline
            static  & 122 & 122 & 121 & 122 \\
            dynamic & 124 & 146 & 50 & 129 \\
            \hline
            clocks  & 79 & 79 & 33 & 66 \\
            signals & 12 & 21 & 6 & 21 \\
            logic   & 10 & 18 & 4 & 17 \\
            %BRAM    & 0 & 4 \\
            DSP     & 23 & 27 & 6 & 25 \\
            \hline
            \hline
        \end{tabular}
    \end{minipage} 
\end{table}

\subsection{Coprocessor/Accelerator Level Results}

In this section, the DLS has been used to derive an accelerator for Xilinx design environments. In particular, according to the possibilities provided by MDC, four different designs have been derived and assessed:

\begin{itemize}
    \item \emph{MM}: memory-mapped accelerator with direct management of data transfer by the host processor;
    \item \emph{MM\_DMA}: memory-mapped accelerator with data transfer managed by a dedicated DMA; \
    \item \emph{STREAM}: stream accelerator with direct management of data transfer by the host processor;
    \item \emph{STREAM\_DMA}: stream accelerator with data transfer managed by a dedicated DMA.
\end{itemize}

Each of the considered accelerators can be configured to execute the DLS algorithm with \emph{BL} or \emph{HP} profile.
%
%\subsection{Results} 
%\textcolor{blue}{FRA: qui secondo me c'è un po di casino con le sezioni, risultati ce n'è anche prima. Avrebbe senso organizzare il tutto in 3? Es: DUT, risultati datapath level, risultati acceleratore? Tiziana: Ho provato a fare questa divisione, date uno sguardo/aggiustate}
%
%In this section results related to the different reconfigurable hardware accelerators generated with MDC are shown. 
Resource occupancy results come from Vivado implementation report of the whole integrated system (involving accelerator, interconnection system and host processor) targeting a Zynq-7000 device (XC7Z020CLG484). Timing results are collected through runs on board of the integrated systems, with the usage of internal host processor timers in order to measure the lasting of the different execution parts.

Table~\ref{tab:acc_res} depicts the resource occupancy of the considered designs, with a detail of the main modules involved in such integrated systems. It is possible to appreciate how the DMA constitutes a quite lightweight additional module for the \emph{MM} system (\emph{MM} versus \emph{MM\_DMA} rows), but it is more resource hungry in the \emph{STREAM} case (\emph{STREAM} versus \emph{STREAM\_DMA} rows). This is because, with \emph{STREAM} communication, one different DMA is required for each couple of I/O ports. Moreover, from Table~\ref{tab:acc_res}, it is clear that the \emph{STREAM} communication is moving resources from the accelerator to the FIFOs and, if enabled, to the DMAs necessary to manage communication between the same accelerator and the host processor. This increase in resource occupancy is justified by an increase in terms of performance, as we are going to show in the following.

\begin{table}
 \begin{center}
 \caption{Resource occupancy of the different integrated systems involving the reconfigurable hardware accelerators generated by MDC. \textcolor{black}{Accelerators have been implemented with an operating frequency of 100 MHz.}}
 \begin{tabular}{cc||cccc}
 \multirow{2}{*}{\textbf{design}} & \multirow{2}{*}{\textbf{module}} & \multicolumn{4}{c}{\textbf{resource}} \\
 & & \textbf{LUT} & \textbf{FF} & \textbf{BRAM} & \textbf{DSP} \\
   \hline
   \hline
  \multirow{2}{*}{\emph{MM}} & accelerator & 21214 & 20646 & 20 & 161 \\
   & total & 22203 & 21868 & 20 & 161 \\
   \hline
  \multirow{3}{*}{\emph{MM\_DMA}} & accelerator & 21183 & 20652 & 20 & 161 \\
   & DMA & 783 & 991 & 0 & 0 \\
   & total & 23951 & 24233 & 20 & 161 \\
   \hline
  \multirow{3}{*}{\emph{STREAM}} & accelerator & 20837 & 20404 & 14 & 161 \\
   & FIFOs & 2128 & 2372 & 6 & 0 \\
   & total & 25792 & 26570 & 20 & 161 \\
   \hline
  \multirow{4}{*}{\emph{STREAM\_DMA}} & accelerator & 20833 & 20404 & 14 & 161 \\
   & FIFOs & 372 & 708 & 9 & 0 \\
   & DMAs & 3586 & 33202 & 29 & 161 \\
   & total & 31798 & 33202 & 29 & 161 \\
 \hline
 \hline
 \end{tabular}
 \label{tab:acc_res}
 \end{center}
 \end{table}

Table~\ref{tab:acc_tim} depicts the execution times of the designs under test, with a detail on the isolated contributions coming from different parts of the accelerators drivers. \textcolor{black}{Such measures have been performed with the accelerators running at their maximum achievable frequency, that is 113.64 MHz, in order to give a fair comparison with a full software execution on the ARM host core, that is instead running at 666.67 MHz.} It is possible to see how, as expected, different execution profiles have similar configuration, loading and storing times, while differ for the processing contribution, where the \emph{HP} profile is overperforming the \emph{BL} one. Going from one design to another, the configuration  and processing time of the two execution profiles are approximately the same, while the other contributions, loading and storing times, differ. In the \emph{MM} cases, these data transfer times are growing when DMA is adopted within the system. Such behavior is not present in the \emph{STREAM} case since memory-mapped to stream conversion is performed even if the DMA is not used, resulting in similar, and rather higher, data transfer times. If MicroBlaze host processor had been chosen, we would have had for the \emph{STREAM} case the same behavior of the \emph{MM} one. Obviously, the \emph{MM} DMA slowdown is not an expected behavior, since the role of DMA is precisely to speed up data transfers and relieve the host processor from the burden of managing them. However, for this specific DLS use case and according to how it has been modeled in dataflow, due to the limited amount of data, 6 data per I/O port at maximum, the DMA management overhead is bigger than the time saved during the data transfer. In summary, for the considered use case, the best choice is opting for \emph{MM} solution without DMA. In order to better understand this point and its relationships with the amount of transmitted data, Table~\ref{tab:acc_load} shows loading times of the considered \emph{MM} accelerators for growing amounts of data to be transferred, from 16 to 256. Here, it is possible to see the benefits of DMA adoption: DMA is capable of saving from 29\% to more than 86\% of the data transfer time.

\begin{table}
 \begin{center}
 \caption{Execution times (tot is total, cfg is configuration, load is input load, proc is processing and store is output store), in terms of the ARM host core clock cycles, of the different integrated systems involving the reconfigurable hardware accelerators generated by MDC under different execution profiles, and of the ARM host core without acceleration. \textcolor{black}{Accelerators are driven with their maximum operating frequency, that is 113.64 MHz.}}
 \begin{tabular}{cc||c|cccc}
 \multirow{2}{*}{\textbf{design}} & \multirow{2}{*}{\textbf{profile}} & \multicolumn{5}{c}{\textbf{time [cck]}} \\
 & & \textbf{tot} & \textbf{cfg} & \textbf{load} & \textbf{proc} & \textbf{store} \\
   \hline
   \hline
  \multirow{2}{*}{\emph{MM}} & \emph{BL} & \textcolor{black}{40527} & 948 & 1352 & \textcolor{black}{37388} & 832 \\
   & \emph{HP} & \textcolor{black}{23062} & 946 & 1360 & \textcolor{black}{19865} & 844 \\
   \hline
  \multirow{2}{*}{\emph{MM\_DMA}} &  \emph{BL} & \textcolor{black}{43309} & 974 & 3454 & \textcolor{black}{37416} & 1544 \\
   & \emph{HP} & \textcolor{black}{25773} & 928 & 3488 & \textcolor{black}{19833} & 1544 \\
   \hline
  \multirow{2}{*}{\emph{STREAM}} & \emph{BL} & \textcolor{black}{41918} & 452 & 3638 & \textcolor{black}{39708} & 1560 \\
   & \emph{HP} & \textcolor{black}{24482} & 394 & 3638 & \textcolor{black}{22137} & 1556 \\
 \hline
  \multirow{2}{*}{\emph{STREAM\_DMA}} & \emph{BL} & \textcolor{black}{40606} & 400 & 3402 & \textcolor{black}{38696} & 1122 \\
   & \emph{HP} & \textcolor{black}{21185} & 404 & 3410 & \textcolor{black}{23136} & 1116 \\
   \hline
   \hline
  \textcolor{black}{\emph{ARM}} & N/A & \textcolor{black}{41204} & - & - & \textcolor{black}{41204} & - \\
   \hline
   \hline
 \end{tabular}
 \label{tab:acc_tim}
 \end{center}
 \end{table}

\begin{comment}
\begin{table}
 \begin{center}
 \caption{Execution times (tot is total, cfg is configuration, load is input load, proc is processing and store is output store), in terms of the ARM host core clock cycles, of the different integrated systems involving the reconfigurable hardware accelerators generated by MDC under different execution profiles. \textcolor{black}{Accelerators are driven with an operating frequency of 100 MHz.}}
 \begin{tabular}{cc||c|cccc}
 \multirow{2}{*}{\textbf{design}} & \multirow{2}{*}{\textbf{profile}} & \multicolumn{5}{c}{\textbf{time [cck]}} \\
 & & \textbf{tot} & \textbf{cfg} & \textbf{load} & \textbf{proc} & \textbf{store} \\
   \hline
   \hline
  \multirow{2}{*}{\emph{MM}} & \emph{BL} & 45594 & 948 & 1352 & 42454 & 832 \\
   & \emph{HP} & 25736 & 946 & 1360 & 22534 & 844 \\
   \hline
  \multirow{2}{*}{\emph{MM\_DMA}} &  \emph{BL} & 48366 & 974 & 3454 & 42482 & 1544 \\
   & \emph{HP} & 28444 & 928 & 3488 & 22502 & 1544 \\
   \hline
  \multirow{2}{*}{\emph{STREAM}} & \emph{BL} & 45354 & 452 & 3638 & 43406 & 1560 \\
   & \emph{HP} & 25444 & 394 & 3638 & 23414 & 1556 \\
 \hline
  \multirow{2}{*}{\emph{STREAM\_DMA}} & \emph{BL} & 45322 & 400 & 3402 & 43762 & 1122 \\
   & \emph{HP} & 25500 & 404 & 3410 & 23854 & 1116 \\
   \hline
   \hline
  \textcolor{black}{\emph{ARM}} & N/A & 41204 & - & - & 41204 & - \\
   \hline
   \hline
 \end{tabular}
 \label{tab:acc_tim}
 \end{center}
 \end{table}
\end{comment}

From Table~\ref{tab:acc_tim} it is also possible to compare \emph{MM} and \emph{STREAM} designs in terms of performance. In particular, the tightly coupled systems (\emph{STREAM} ones) are better than \emph{MM} ones either when DMA is adopted, while without DMA such difference is not appreciable. Thanks to the different communication protocol directly connected with the reconfigurable computing core, it is possible to overlap loading/storing data and processing in the \emph{STREAM} case. While data is being sent to one input port of the reconfigurable computing core, the previously sent data is already under processing; and while data is being processed, data that have been already processed is under sending from one output port of the reconfigurable computing core. The advantage provided by the \emph{STREAM} designs is however only barely visible and only in the DMA case, again depending on the characteristics of the accelerated application. In fact, in front of a very small amount of I/O data, there is a huge amount of processing, as can be seen by comparing load or store columns with proc ones in Table~\ref{tab:acc_tim}. Thus, the overlapping between data transfer and processing, that is directly proportional to the advantages of the \emph{STREAM} designs, is small.

\textcolor{black}{Lastly, Table~\ref{tab:acc_tim} is also giving an idea of the acceleration capabilities of the developed systems, if the \emph{ARM} row is considered. \emph{ARM} row reports on the clock cycles needed by the ARM host processor to run the DLS application. Here, the source code is the same adopted for the actors synthesis through Vivado HLS. Of course, being the execution profiles the result of the same Vivado HLS optimizations, in the ARM full software execution there are not different execution profiles. In terms of speed-up, the \emph{BL} profile over the different accelerators uses more or less the same amount of cycles than the ARM %the developed accelerators with BL profile are more or less employing the same time of the ARM core to terminate the application,
meaning that no speed-up is achieved. The reason to adopt them would be just to %Thus, in the BL case, accelerators can be only used to 
relieve the ARM core from the burden of executing the trajectory computation, letting it free of performing other tasks. However, when the \emph{HP} profile is enabled, accelerators deliver a speed-up that is close to 2x. Please consider that, for the purposes of this work, a limited effort has been put in optimizing the acceleration itself, since demonstrating MDC acceleration capabilities is not among the contributions of the paper. Better exploiting the capabilities of Vivado HLS (e.g. by adopting more pragmas or by deriving more high performance actors), working on the model (e.g. by parallelizing the critical actors), and taking advantage of the other features of MDC tool (e.g. the dynamic power manager), the user could achieve even better results and wider trade-offs under all the considered metrics.}

\begin{table}
 \begin{center}
 \caption{Detail of the input load times of the different integrated systems involving the reconfigurable hardware accelerators generated by MDC for different data amount.}
 \begin{tabular}{cc||c|c}
 \textbf{design} & \textbf{data amount} & \textbf{time [cck]} & \textbf{\% vs no DMA} \\
   \hline
   \hline
  \multirow{5}{*}{\emph{MM}} & 16 & 1880 & - \\
   & 32 & 3612 & - \\
   & 64 & 7132 & - \\
   & 128 & 14170 & - \\
   & 256 & 28254 & - \\
   \hline
   \hline
  \multirow{5}{*}{\emph{MM\_DMA}} &  16 & 1344 & -28.51 \\
   & 32 & 1492 & -58.69 \\
   & 64 & 1760 & -75.32 \\
   & 128 & 2468 & -82.58 \\
   & 256 & 3854 & -86.36 \\
   \hline
   \hline
  %\multirow{5}{*}{\emph{STREAM}} & 16 &  &  \\
  % & 32 &  &  \\
  % & 64 &  &  \\
  % & 128 &  &  \\
  % & 256 &  &  \\
  % \hline
  % \hline
  %\multirow{5}{*}{\emph{STREAM\_DMA}} & 16 &  &  \\
  % & 32 &  &  \\
  % & 64 &  &  \\
  % & 128 &  &  \\
  % & 256 &  &  \\
 %\hline
 %\hline
 \end{tabular}
 \label{tab:acc_load}
 \end{center}
 \end{table}

\section{Usage of MDC Tool}
\label{s:usage}

\textcolor{black}{
It is worth to make some considerations also on the usage of MDC in terms of design time and effort. It has been  already discussed that the design of MDC compliant CGR systems is not straightforward, it is time consuming and error prone. MDC compliant system composition is application specific and the interconnection infrastructure is irregular, since MDC does not leverage onto an homogeneous CGR array. The designer should analyze the input networks, identify the common resources in the different dataflow specifications, and combine them, keeping trace of the actors that belong to different functionalities to program the multiplexers properly. %The usage of 
MDC speeds-up and simplifies the design of the CGR datapath, by automatically mapping different input specifications in one single reconfigurable substrate, addressing all the above mentioned steps. Nevertheless, the usage of that substrate requires additional steps. In fact, users should be capable to pack it in a coprocessor with its own APIs, as we have seen to access it as a computational resource. Also this system integration step is time consuming and error prone. Therefore, having a tool capable of going from the generated CGR datapath to the integrated processor-coprocessor system, with user friendly drivers to be inserted in the host code, is certainly beneficial for any potential MDC user.}%Also this latter point has to be highlighted, since a time consuming and error prone step is also constituted by system integration activities, in this case going from the generated CGR datapath to the integrated processor-coprocessor system with user friendly drivers to be inserted in the host code.}

\textcolor{black}{
The effort of designing the dataflow specifications is application specific and designer specific. This is true for any kind of coding, including imperative one, since according to the designer skills and to the complexity of the application the required time could vary a lot, resulting in a hardly quantifiable metric. The usage of dataflow MoCs forces designers to think in a modular way favoring code re-use, which has a positive impact on time to deployment, and not preventing the adoption of the more common imperative code to describe the functionality embedded by each actor. In fact, %from the beginning, %by embedding imperative code within actors that are concurrently executed according to data availability. Despite such change of setting, the development of dataflow specifications could then be not so far from the one of common imperative specifications. For instance, 
actors can be specified with simple C code, leveraging on standard HLS engines, such as Vivado HLS, for their translation onto HDL, while designers have only to take care about dataflow network specification, as additional tasks. As described in Section \ref{ss:tool_dataflow}, many dataflow-based tools are available, from optimization and mapping (e.g. PREESM~\cite{PREESM}) to HLS ones (e.g. CAPH~\cite{Serot_2013}). Their combined usage represents for sure a benefit to solve many different design issues. %, on one side speed the design process up by providing design automation for several development step, and on the other side let software developers, that might not be expert of hardware design, access the proposed design flow.
}

\textcolor{black}{
Going back to MDC, it is certainly true that users need to specify the applications through abstract high-level input dataflow representations. Nevertheless, having accomplished that, the toolchain takes care of the complete process, generating all the necessary files to implement a processor-coprocessor system.} \textcolor{black}{In particular, the time necessary for all the MDC steps including parsing input dataflow specifications, merging them, and generating the output files is in the order of seconds. While the system deployment, which involves opening Vivado and running the TCL files generated by MDC, requires about one minute.\footnote{Timing has been estimated on a laptop with an Intel(R) Core(TM) i7-4500U CPU @ 1.80GHz, with 8 GB RAM, running Ubuntu 16.04.}} \textcolor{black}{Considering the remaining implementation steps, everything is more or less automated and designers can leverage on commercial tools, but the requested design time is again application and target dependent. Indeed the time necessary to synthesize and implement the design, and generate the bitstream, depends on the size of the target device and on the size of the application to be mapped over it.}

\section{MDC in a Bigger Picture}
\label{sect:interf}
%\subsection{MDC Interfaced with other Tools}\label{sect:interf}
Designing CPS requires acting at different levels of the system, that can be identified as: 1) Application level, 2) Run-time Management level, and 3) Architectural level.
MDC spans across all of those levels, offering the possibility of designing the application to be accelerated as dataflow specification (application level), generating automatically the corresponding multi-functional accelerator (Architectural level) and providing the necessary drivers and APIs to manage the accelerator (Run-time Management level). Currently, MDC is an open-source tool available on GitHub.\footnote{https://github.com/mdc-suite/mdc}.

Many other tools for the design of CPS systems, and their sub-parts, there exist. They work at the same or at different level of the proposed design flow and, some of them, are compliant and complementary to MDC functionalities. Along the years, a huge effort has been put in integrating MDC with some of these tools to extend MDC applicability and features. The following sections briefly describe how MDC is currently interfaced with other tools.

\subsection{Exploiting High-Level Synthesis}\label{ss:caph}
MDC requires the HDL hardware descriptions of the dataflow actors. Deriving it directly from the dataflow models by means of HLS engines could be convenient. In literature, HLS is a hot topic and several HLS engines have been proposed either from academy (e.g. Xronos \cite{Bezati_2013_ESL}, CAPH \cite{Serot_2013}, Bambu \cite{bambu}) and industry (e.g. Vivado HLS \cite{vivado_hls}, Intel FPGA SDK for OpenCL \cite{intel_sdk}, Cadence Stratus \cite{cadence_stratus}). 
In the past, MDC was interfaced with the Xronos HLS engine, being based on the same dataflow MoC. In spite of benefits in terms of design time, Xronos adoption lead to a strong limitation: the target platforms were fixed the ones of one vendor: Xilinx FPGAs. In the CPS context, in which a support for a wide range of systems is required, this limitation led to the MDC hardware communication protocol generalization and to the CAPH-MDC integration.
Generally speaking, there is no perfect HLS engine. The efficiency of the obtained systems is linked to the context of application, and highly depends on the target device/technology, as well as on the initial specification format \cite{Nane2016}. A novel choice in this sense is CAPH, an open source target independent HLS engine supporting dataflow models as specification formats, similar to the MDC ones. CAPH generates generic RTL descriptions for any kind of FPGA or even for ASIC flows. 
Thus, MDC has been integrated with CAPH to provide a generic fully automated CGR flow \cite{Rubattu_2018_Embedded}. %The aim of this process is not to add another HLS engine, besides Xronos, among the MDC supported ones, but to allow the designer to choose any kind of HLS engine. 
This goal required two main actions:
\begin{itemize}
    \item a CAPH-to-XDF parser has been defined to implement model-to-model transformations from CAPH dataflows to MDC compliant dataflows \cite{Bhattacharyya_2011}.
    \item a generalization of the supported actor-to-actor communication protocol, originally fixed and compliant with MPEG-RVC actors only, to support in hardware any user-defined actor-to-actor communication handshake.
\end{itemize}
%Besides customizing the communication handshake, users have now additional features: they can link model parameters to the HDL specification, use additional modules as actors (such as FIFOs and fanouts), and specify system level signals (e.g. clock and reset). With the MDC \& CAPH integration, it is possible to automatically generate generic CG reconfigurable accelerators for the CERBERO adaptivity support. Such kind of reconfigurability can be also reached by means of imperative (non dataflow oriented) HLS engines, such as Vivado HLS or Altera HLS Compiler. 
Please note that the second bullet allows MDC to be combined with potentially any HLS tool, also with imperative (non dataflow oriented) HLS engines, such as Vivado HLS, Intel FPGA SDK for OpenCL or Cadence Stratus. %Nevertheless, it is worth to be mentioned that,} with respect to imperative HLS, the integration of MDC with CAPH led to unprecedented predictability for performance indicators such as latency and throughput, before the synthesis stage. By contrast, Vivado HLS and Altera HLS Compiler tools provide latency only after synthesis and only for simple designs: if reconfiguration is implemented on the top of the imperative language, latency estimation may not be accurate or even available.

\subsection{High-Level Profiling and Run-time Management}\label{ss:spider}
In order to exploit the hardware acceleration provided by MDC in a higher software-level application, a design flow that combines PREESM, SPIDER and MDC has been derived. PREESM is a tool capable of scheduling and mapping dataflow applications onto multi- and many-core architectures \cite{PREESM}. While SPIDER is the run-time simplified version of PREESM, providing software scheduling and memory management during execution \cite{Heulot2014}. The motivation at the base of the tool combination was the possibility of offering software reconfigurability management of PREESM and SPIDER with hardware reconfigurability management of MDC, basing on their complementary characteristics. 

At design time, the proposed integration implies the creation of the application graph, conform to a dataflow MoC, and a software task in which the processing is delegated to the accelerator by using the driver functions available after the MDC coprocessor generation. This task corresponds to the high-level dataflow actor that has to be accelerated on hardware. In the run-time context, SPIDER is capable of configuring the CGR accelerators generated by MDC to compute their different functionalities, which can be selected dynamically through dynamic parameters. Depending on the adaptation strategy, SPIDER schedules and maps, at run-time, the whole high-level application graph composed of software tasks, including those that manage the communication with the accelerator, and sends these latter to the slave processors, as described in \cite{Rubattu_2018_CPSWS}. 

\subsection{Enabling feedback for self-adaptation: HW/SW Monitoring}\label{ss:papi}
When the CPS is requested to be not just adaptive, but self-adaptive, it is crucial to enable a feedback to communicate the system status to a run-time manager. To enable such a feedback on the system status, a monitoring infrastructure able to read both the monitors normally available on standard CPUs and custom monitors that may be inserted on the hardware accelerator is necessary. For this reason MDC has been integrated with \textsc{Papify}, to provide a toolchain able to offer the support in the process of designing, implementing and managing monitored CGR substrates~\cite{FanniT_2019}.

Placed between Run-time Management and Architectural levels, \textsc{Papify} tool generalizes PAPI\footnote{PAPI provides a unified method to access the PMCs available on the CPUs~\cite{papi}.} for embedded heterogeneous architectures~\cite{Madronal_2019_Access}, offering an interface to access the performance monitoring information of the different PEs existing in the target platform. \textcolor{black}{The system generation capabilities of MDC have been extended to offer the possibility of including accelerator level monitors, which are PAPI-compliant and can be accessed through \textsc{Papify}. When users enable the system generation, they only need to ticks the monitor related boxes in MDC GUI to also enable the automatic instrumentation of the design.}

The users need to specify the applications as dataflow specifications, as for the common MDC features, then the complete process from dataflow to the processor-coprocessor system is automatically carried out by the tool. The APIs generated by MDC mask the complexity of the processor-coprocessor communication, and the support for heterogeneous architectures provided by \textsc{Papify} masks the access to the monitors. 

\subsection{New Level of flexibility: the multigrain reconfiguration}\label{ss:artico}
Reconfigurable hardware architectures present high performance and flexibility, being an appealing solution to provide run-time adaptivity support necessary for CPS. MDC offers the CGR approach for delivering such architectures. Another approach, highly adopted at this purpose, is the DPR one.  CGR has lower overhead than DPR but it is in general less flexible. The combination of these two hardware reconfiguration approaches brings together the best of both, offering the possibility of achieving different trade-offs between performance, flexibility and energy consumption. 

Placed between Run-time Management and Architectural levels, the ARTICo\textsuperscript{3} framework provides adaptive and scalable hardware acceleration by exploiting a DPR-based multi-accelerator scheme, leveraging on reconfigurable slots~\cite{Rodriguez_2018}. Furthermore, it provides a \textit{Run-time Library} to manage the application execution and computation offloading to the hardware accelerators, i.e. to the slots.

\textcolor{black}{The system generation capabilities of MDC have been extended with the addition of a new backend, that generates CGR substrates compliant with ARTICo\textsuperscript{3} slots.}
The integrated MDC-ARTICo\textsuperscript{3} toolchain offers a new level of flexibility, combining together CGR and DPR. The toolchain maps different input specifications in one CGR datapath compliant with the DPR-based ARTICo\textsuperscript{3} slots, speeding-up the design of multi-grain systems. Users only need to define the applications behavior through abstract high-level input dataflow specifications. The management of the generated multi-grain system is on the \textit{Run-time Library} of the ARTICo\textsuperscript{3} architecture that is naturally capable of managing hardware accelerators also when these latter are CGR substrates \cite{Fanni_2018}.%,Rodriguez_2018x2}.

\section{Conclusions}\label{s:concl}

The Multi-Dataflow Composer tool has been successfully designed and assessed in the context of signal processing systems. In particular it demonstrated, along the years, to be applicable to the video coding field \cite{Sau_2017}, and more in general, to the needs for flexibility of cyber-physical systems \cite{PalumboSFR17}. %Thanks to its usage in different projects, as the H2020 CERBERO project\footnote{https://www.cerbero-h2020.eu/} \cite{Masin_2017,Palumbo:2019CF} and the ECSEL FitOptiVis project\footnote{http://fitoptivis.utu.fi/} \cite{zaid_2019}, 
\textcolor{black}{Thanks to its usage in different projects\footnote{https://www.cerbero-h2020.eu/}$^{,}$\footnote{http://fitoptivis.utu.fi/} \cite{Palumbo:2019CF, zaid_2019}}
MDC has been extended with new supports and functionalities to make hardware accelerators and heterogeneous systems more easy to be developed and used by software programmers, by increasing the level of abstraction and avoiding them to face the daunting tasks of handling low level details of datapath generation, system customization and optimization. 

At the moment, a particular effort is put in improving and optimizing system generation, by making it faster leveraging on mathematical programming and algebraic optimization strategies, and on relieving the users also from an additional burden, which is system partitioning. In this latter case, we are studying ways to model this kind of architectures in order to allow not only for dataflow-based hardware/software partitioning, but also advanced dynamic re-mapping and reconfiguration. In terms of engineering effort, the \emph{Coprocessor Generator}, currently supporting Xilinx FPGA environments only, in future will be extended to target Intel FPGA ones, as well as generic ASIC platforms. Last, but not least, MDC has been in the last years on its way to open-source, and different technology transfer activities\footnote{As the technology transfer activities carried out within the Sardinian Regional project PROSSIMO: www.cluster-prossimo.it} helped in improving its usability and in defining a concrete path to the market. 
\appendix
\section{Listings}
\label{app:list}

\lstset{language=C,caption={Coprocessor drivers interface.}, commentstyle=\color{mygreen}, keywordstyle=\color{mymauve}, label={list:int_dr},basicstyle=\footnotesize}
\begin{lstlisting}
/////////////////////////////////
//Memory-Mapped Interface Driver
int mm_accelerator_roberts(
// port out_pel
int size_out_pel, int* data_out_pel,
// port in_pel
int size_in_pel, int* data_in_pel,
// port in_size
int size_in_size, int* data_in_size
)
...

/////////////////////////////////
// Stream-Based Interface Driver
int s_accelerator_roberts(
// port out_pel
int size_out_pel, int* data_out_pel,
// port in_pel
int size_in_pel, int* data_in_pel,
// port in_size
int size_in_size, int* data_in_size
) 
...
\end{lstlisting}

\lstset{language=C,caption={Coprocessor drivers body.}, commentstyle=\color{mygreen},  keywordstyle=\color{mymauve}, label={list:bodyDr},basicstyle=\footnotesize}
\begin{lstlisting}
/////////////////////////////
// Memory-Mapped Body Driver
...
// configure I/O
*(config + 1) = size_in_size;
...
// send data port in_size
*((volatile int*) XPAR_AXI_CDMA_0_BASEADDR + (0x04>>2)) = 0x00000002; // verify idle
*((volatile int*) XPAR_AXI_CDMA_0_BASEADDR + (0x18>>2)) = (int) data_in_size; // src
*((volatile int*) XPAR_AXI_CDMA_0_BASEADDR + (0x20>>2)) = 
XPAR_MM_ACCELERATOR_0_MEM_BASEADDR + MM_ACCELERATOR_MEM_1_OFFSET; // dst
*((volatile int*) XPAR_AXI_CDMA_0_BASEADDR + (0x28>>2)) = size_in_size*4; // size [B]
while((*((volatile int*) XPAR_AXI_CDMA_0_BASEADDR + (0x04>>2)) & 0x2) != 0x2);
...

// receive data port out_pel
*((volatile int*) XPAR_AXI_CDMA_0_BASEADDR + (0x04>>2)) = 0x00000002; // verify idle
*((volatile int*) XPAR_AXI_CDMA_0_BASEADDR + (0x18>>2)) = 
XPAR_MM_ACCELERATOR_0_MEM_BASEADDR + MM_ACCELERATOR_MEM_3_OFFSET; // src
*((volatile int*) XPAR_AXI_CDMA_0_BASEADDR + (0x20>>2)) = (int) data_out_pel; // dst
*((volatile int*) XPAR_AXI_CDMA_0_BASEADDR + (0x28>>2)) = size_out_pel*4; // size [B]
while((*((volatile int*) XPAR_AXI_CDMA_0_BASEADDR + (0x04>>2)) & 0x2) != 0x2);
...

////////////////////////////
// Stream-Based Body Driver
...
// configure I/O
*((int*) (XPAR_S_ACCELERATOR_0_CFG_BASEADDR + 1*4)) = size_out_pel;

// start execution
*(config) = 0x2000001;

// send data port in_size
*((volatile int*) XPAR_AXI_DMA_0_BASEADDR + (0x00>>2)) = 0x00000001; // start
*((volatile int*) XPAR_AXI_DMA_0_BASEADDR + (0x04>>2)) = 0x00000000; // reset idle
*((volatile int*) XPAR_AXI_DMA_0_BASEADDR + (0x18>>2)) = (int) data_in_size; // src
*((volatile int*) XPAR_AXI_DMA_0_BASEADDR + (0x28>>2)) = size_in_size*4; // size [B]
while(((*((volatile int*) XPAR_AXI_DMA_0_BASEADDR + (0x04>>2))) & 0x2) != 0x2);
...

// receive data port out_pel
*((volatile int*) XPAR_AXI_DMA_0_BASEADDR + (0x30>>2)) = 0x00000001; // start
*((volatile int*) XPAR_AXI_DMA_0_BASEADDR + (0x34>>2)) = 0x00000000; // reset idle
*((volatile int*) XPAR_AXI_DMA_0_BASEADDR + (0x48>>2)) = (int) data_out_pel; // dst
*((volatile int*) XPAR_AXI_DMA_0_BASEADDR + (0x58>>2)) = size_out_pel*4; // size [B]
while(((*((volatile int*) XPAR_AXI_DMA_0_BASEADDR + (0x34>>2))) & 0x2) != 0x2);
...
\end{lstlisting}

\lstset{language=tcl,caption={IP Generation Script.},label={list:ip_scr}, commentstyle=\color{mygreen}, basicstyle=\footnotesize}
\begin{lstlisting}
###########################
# IP Settings
###########################

...
# FPGA device
set partname "xc7z020clg400-1"
set boardpart "digilentinc.com:arty-z7-20:part0:1.0"

# Design name
set ip_name "mm_accelerator"
set design $ip_name

###########################
# Create IP
###########################

create_project -force $design $ipdir -part $partname 
set_property board_part $boardpart [current_project]
set_property target_language Verilog [current_project]

add_files $hdl_files_path
import_files -force

set files [glob -tails -directory $ipdir/.../lib/caph/ *]
foreach f $files {
set name $f
set_property library caph [get_files  $ipdir/.../lib/caph/$f]
}

set_property top $ip_name [current_fileset]

ipx::package_project -root_dir $ipdir -vendor user.org\
-library user -taxonomy AXI_Peripheral

ipx::add_address_block s00_axi_reg\
[ipx::get_memory_maps s00_axi -of_objects [ipx::current_core]]
ipx::add_address_block s01_axi_mem\
[ipx::get_memory_maps s01_axi -of_objects [ipx::current_core]]
...

file copy -force $iproot/drivers $ipdir
set drivers_dir drivers
ipx::add_file_group -type software_driver {} [ipx::current_core]
...

set_property core_revision 3 [ipx::current_core]
ipx::create_xgui_files [ipx::current_core]
ipx::update_checksums [ipx::current_core]
ipx::save_core [ipx::current_core]
set_property  ip_repo_paths $ipdir [current_project]
update_ip_catalog
close_project
\end{lstlisting}

\lstset{language=tcl,caption={Top Design Generation Script}, commentstyle=\color{mygreen}, label={list:top_scr},basicstyle=\footnotesize}
\begin{lstlisting}
###########################
# Settings
###########################

...
# FPGA device
set partname "xc7z020clg400-1"
set boardpart "digilentinc.com:arty-z7-20:part0:1.0"

# Design name
set design system
set bd_design "design_1"

###########################
# Create Project
###########################
create_project -force $design $projdir -part $partname 
set_property board_part $boardpart [current_project]
set_property target_language Verilog [current_project]
set_property  ip_repo_paths $ipdir [current_project]
update_ip_catalog -rebuild -scan_changes
###########################
#create block design
create_bd_design $bd_design

# Zynq PS
create_bd_cell -type ip\
-vlnv xilinx.com:ip:processing_system7:5.5 processing_system7_0
...

# accelerator IP
create_bd_cell -type ip -vlnv user.org:user:$ip_name:$ip_version $ip_name\_0

apply_bd_automation -rule xilinx.com:bd_rule:axi4\
-config {...}  [get_bd_intf_pins $ip_name\_0/s00_axi]

# CDMA
create_bd_cell -type ip -vlnv xilinx.com:ip:axi_cdma:4.1 axi_cdma_0
set_property -dict [list CONFIG.C_INCLUDE_SG {0}] [get_bd_cells axi_cdma_0]

apply_bd_automation -rule xilinx.com:bd_rule:axi4\
-config {...}  [get_bd_intf_pins axi_cdma_0/S_AXI_LITE]
...

make_wrapper -files [get_files $projdir/.../design_1.bd] -top
add_files -norecurse $projdir/.../hdl/design_1_wrapper.v
...
\end{lstlisting}

\section*{List of Acronyms}
\begin{itemize}
    \item ASIC: Application Specific Integrated Circuit
    \item API: Application Program Interfaces
    \item BL: Baseline
    \item CAL: Caltrop Actor Language
    \item CGR: Coarse-Grain Reconfigurable
    \item CP: Critical Path
    \item CPF: Common Power Format
    \item CPS: Cyber-Physical Systems
    \item CPU: Central Processing Unit
    \item DFG: Data-Flow Graph
    \item DLS: Damped  Least  Square
    \item DMA: Direct Memory Access
    \item DPN: Dataflow Processing Network
    \item DPR: Dynamic and Partial Reconfiguration
    \item DSP: Digital Signal Processing
    \item F: functional
    \item FF: Flip-Flop
    \item FIFO: First-In First-Out 
    \item FPGA: Field Programmable Gate Array
    \item GUI: Graphical User Interface
    \item HDL: Hardware Description Language
    \item HLS: High Level Synthesis
    \item HP: High Performance
    \item IK: Inverse Kinematics
    \item I/O Input/Output
    \item IP: Intellectual Property
    \item IR: Intermediate Representation
    \item LR: Logic Region
    \item LUT:  Look-Up Table
    \item LWDF: Lightweight dataflow
    \item MDC: Multi-Dataflow Copmposer
    \item mm/MM: memory mapped
    \item MoC: Model of Computation
    \item MPEG-RVC: MPEG Reconfigurable Video Coding
    \item NF: Non-Functional
    \item ORCC: Open RVC-CAL Compiler
    \item PD: Power Domain
    \item PE: Processing Element
    \item PiSDF: Parameterized and Interfaced Synchronous Dataflow
    \item RTL: Register Transfer Level
    \item SBox: Switching Box
    \item s: stream
    \item TIL: Template Interface Layer
    \item UPF: Unified Power Format
    \item XDF:  XML Dataflow Format
\end{itemize}

\section*{Acknowledgments}
Authors would like to thank Eng. Luca Fanni, a former employ of the University of Sassari, for providing the initial MATLAB code used to derived the presented use case. Dr. Francesca Palumbo is grateful to the University of Sassari that supported her studies on MDC related activities through the ``fondo di Ateneo per la ricerca 2019'' and to the Comp4Drones project (No 826610, ECSEL-JU 2018) that will continue to found them in the next years. Moreover, all the MDC activities in both involved universities have been carried out so far \textcolor{black}{as part of the FitOptiVis project \cite{zaid_2019}, funded by the ECSEL Joint Undertaking under grant number H2020-ECSEL-2017-2-783162, and of the CERBERO H2020 project \cite{Masin_2017,Palumbo:2019CF}, funded by European Union under grant number No 732105}. 
MDC is also part of technology transfer activities carried out by the University of Sassari within the the PROSSIMO project (POR FESR 2014/20-ASSE I), for which the authors would like to thanks the Sardinian Regional Government. 

%\section*{References}

%\bibliographystyle{elsarticle-num}
%\bibliography{elsarticle-template}

\begin{thebibliography}{10}
\expandafter\ifx\csname url\endcsname\relax
  \def\url#1{\texttt{#1}}\fi
\expandafter\ifx\csname urlprefix\endcsname\relax\def\urlprefix{URL }\fi
\expandafter\ifx\csname href\endcsname\relax
  \def\href#1#2{#2} \def\path#1{#1}\fi

\bibitem{Xilinx}
Xilinx.
\newblock \href{https://www.xilinx.com}{[link]}.
\newline\urlprefix\url{https://www.xilinx.com}

\bibitem{Intel}
Intel.
\newblock \href{https://www.intel.com}{[link]}.
\newline\urlprefix\url{https://www.intel.com}

\bibitem{Cadence}
Cadence.
\newblock \href{https://www.cadence.com}{[link]}.
\newline\urlprefix\url{https://www.cadence.com}

\bibitem{Compton_2002}
K.~Compton, S.~Hauck, Reconfigurable computing: A survey of systems and
  software, ACM Computing Surveys 34~(2) (2002) 171--210.
\newblock \href {http://dx.doi.org/10.1145/508352.508353}
  {\path{doi:10.1145/508352.508353}}.

\bibitem{Altera:partial_reconf}
Altera,
  \href{https://www.intel.com/content/dam/www/programmable/us/en/pdfs/literature/wp/wp-01137-stxv-dynamic-partial-reconfig.pdf}{Increasing
  Design Functionality with Partial and Dynamic Reconfiguration in 28-nm FPGAs}
  (2010).
\newline\urlprefix\url{https://www.intel.com/content/dam/www/programmable/us/en/pdfs/literature/wp/wp-01137-stxv-dynamic-partial-reconfig.pdf}

\bibitem{Xilinx:partial_reconf}
Xilinx,
  \href{https://www.xilinx.com/support/documentation/sw_manuals/xilinx14_7/ug702.pdf}{Partial
  Reconfiguration User Guide} (April 2013).
\newline\urlprefix\url{https://www.xilinx.com/support/documentation/sw_manuals/xilinx14_7/ug702.pdf}

\bibitem{PalumboFSRMDMPR19}
F.~Palumbo, T.~Fanni, C.~Sau, A.~Rodr{\'{\i}}guez, D.~Madro{\~{n}}al,
  K.~Desnos, A.~Morvan, M.~Pelcat, C.~Rubattu, R.~Lazcano, L.~Raffo, E.~de~la
  Torre, E.~Ju{\'{a}}rez, C.~Sanz, P.~S. de~Rojas, Hardware/software
  self-adaptation in {CPS:} the {CERBERO} project approach, in: International
  Conference on Embedded Computer Systems: Architectures, Modeling, and
  Simulation ({SAMOS}), 2019, pp. 416--428.
\newblock \href {http://dx.doi.org/10.1007/978-3-030-27562-4_30}
  {\path{doi:10.1007/978-3-030-27562-4_30}}.

\bibitem{PalumboSFR17}
F.~Palumbo, C.~Sau, T.~Fanni, L.~Raffo, Challenging {CPS} trade-off adaptivity
  with coarse-grained reconfiguration, in: Applications in Electronics
  Pervading Industry, Environment and Society ({APPLEPIES}), 2017, pp. 57--63.
\newblock \href {http://dx.doi.org/10.1007/978-3-319-93082-4\_8}
  {\path{doi:10.1007/978-3-319-93082-4\_8}}.

\bibitem{Yan_2012}
M.~Yan, Z.~Yang, L.~Liu, S.~Li, Prodfa: Accelerating domain applications with a
  coarse-grained runtime reconfigurable architecture, in: 2012 IEEE 18th
  International Conference on Parallel and Distributed Systems (ICPADS), 2012,
  pp. 834--839.
\newblock \href {http://dx.doi.org/10.1109/ICPADS.2012.136}
  {\path{doi:10.1109/ICPADS.2012.136}}.

\bibitem{Palumbo_2016}
F.~Palumbo, T.~Fanni, C.~Sau, P.~Meloni, Power-awarness in coarse-grained
  reconfigurable multi-functional architectures: a dataflow based strategy,
  Journal of Signal Processing Systems (2016) 1--26\href
  {http://dx.doi.org/10.1007/s11265-016-1106-9}
  {\path{doi:10.1007/s11265-016-1106-9}}.

\bibitem{Ansaloni_2012}
G.~Ansaloni, K.~Tanimura, L.~Pozzi, N.~Dutt, Integrated kernel partitioning and
  scheduling for coarse-grained reconfigurable arrays, IEEE Transactions on
  Computer-Aided Design of Integrated Circuits and Systems 31~(12) (2012)
  1803--1816.
\newblock \href {http://dx.doi.org/10.1109/TCAD.2012.2209886}
  {\path{doi:10.1109/TCAD.2012.2209886}}.

\bibitem{Sau_2015}
C.~Sau, L.~Fanni, P.~Meloni, L.~Raffo, F.~Palumbo, Reconfigurable coprocessors
  synthesis in the {MPEG-RVC} domain, in: International Conference on
  ReConFigurable Computing and FPGAs (ReConFig), 2015, pp. 1--8.
\newblock \href {http://dx.doi.org/10.1109/ReConFig.2015.7393351}
  {\path{doi:10.1109/ReConFig.2015.7393351}}.

\bibitem{Tessier_2001}
R.~Tessier, W.~Burleson, Reconfigurable computing for digital signal
  processing: A survey, Journal of Signal Processing Systems 28~(1-2) (2001)
  7--27.
\newblock \href {http://dx.doi.org/10.1023/A:1008155020711}
  {\path{doi:10.1023/A:1008155020711}}.

\bibitem{Todman_2005}
T.~Todman, G.~Constantinides, S.~Wilton, O.~Mencer, W.~Luk, P.~Cheung,
  Reconfigurable computing: architectures and design methods, IEE
  Proceedings-Computers and Digital Techniques 152~(2) (2005) 193--207.
\newblock \href {http://dx.doi.org/10.1049/ip-cdt:20045086}
  {\path{doi:10.1049/ip-cdt:20045086}}.

\bibitem{Bhat_2013x1}
S.~S. Bhattacharyya, E.~Deprettere, R.~Leupers, J.~Takala (Eds.), Handbook of
  Signal Processing Systems, 2nd Edition, Springer, 2013, iSBN:
  978-1-4614-6858-5 (Print); 978-1-4614-6859-2 (Online).
\newblock \href {http://dx.doi.org/10.1007/978-1-4614-6859-2}
  {\path{doi:10.1007/978-1-4614-6859-2}}.

\bibitem{Dennis_1974}
J.~B. Dennis, First version of a data flow procedure language, in: Programming
  Symposium, Proceedings Colloque Sur La Programmation, Springer-Verlag, 1974,
  pp. 362--376.
\newblock \href {http://dx.doi.org/10.1007/3-540-06859-7_145}
  {\path{doi:10.1007/3-540-06859-7_145}}.

\bibitem{Kahn_1974}
K.~Gilles, The semantics of a simple language for parallel programming, In
  Information Processing 74 (1974) 471--475.

\bibitem{Lee_1995}
E.~Lee, T.~Parks, Dataflow process networks, Proceedings of the IEEE 83~(5)
  (1995) 773--801.
\newblock \href {http://dx.doi.org/10.1109/5.381846}
  {\path{doi:10.1109/5.381846}}.

\bibitem{McAllister_2004}
J.~McAllister, R.~Woods, R.~Walke, D.~Reilly, Synthesis and high level
  optimisation of multidimensional dataflow actor networks on {FPGA}, in:
  Proceedings of the IEEE Workshop on Signal Processing Systems (SIPS), 2004.
\newblock \href {http://dx.doi.org/10.1109/SIPS.2004.1363043}
  {\path{doi:10.1109/SIPS.2004.1363043}}.

\bibitem{Stefanov_2004}
T.~Stefanov, C.~Zissulescu, A.~Turjan, B.~Kienhuis, E.~Deprettere, System
  design using {Kahn} process networks: the {Compaan}/{Laura} approach, in:
  Proceedings of the Design, Automation and Test in Europe Conference and
  Exhibition (DATE), 2004.
\newblock \href {http://dx.doi.org/10.1109/DATE.2004.1268870}
  {\path{doi:10.1109/DATE.2004.1268870}}.

\bibitem{Pelcat_2014}
M.~Pelcat, K.~Desnos, J.~Heulot, C.~Guy, J.~Nezan, S.~Aridhi, Preesm: A
  dataflow-based rapid prototyping framework for simplifying multicore dsp
  programming, in: 2014 6th European Embedded Design in Education and Research
  Conference (EDERC), 2014, pp. 36--40.
\newblock \href {http://dx.doi.org/10.1109/EDERC.2014.6924354}
  {\path{doi:10.1109/EDERC.2014.6924354}}.

\bibitem{Desnos_2013}
K.~{Desnos}, M.~{Pelcat}, J.~{Nezan}, S.~S. {Bhattacharyya}, S.~{Aridhi}, Pimm:
  {Parameterized} and interfaced dataflow meta-model for mpsocs runtime
  reconfiguration, in: {International} {Conference} on {Embedded} {Computer}
  {Systems}: {Architectures}, {Modeling}, and {Simulation} ({SAMOS}), 2013.
\newblock \href {http://dx.doi.org/10.1109/SAMOS.2013.6621104}
  {\path{doi:10.1109/SAMOS.2013.6621104}}.

\bibitem{Orcc}
{RVC-CAL Community}, \href{http://orcc.sourceforge.net/}{{O}pen {RVC}-{CAL}
  compiler ({O}rcc)} (2018).
\newline\urlprefix\url{http://orcc.sourceforge.net/}

\bibitem{Siret_2010}
N.~Siret, I.~Sabry, J.~Nezan, M.~Raulet, A codesign synthesis from an mpeg-4
  decoder dataflow description, in: Proceedings of 2010 IEEE International
  Symposium on Circuits and Systems (ISCAS), 2010, pp. 1995--1998.
\newblock \href {http://dx.doi.org/10.1109/ISCAS.2010.5537107}
  {\path{doi:10.1109/ISCAS.2010.5537107}}.

\bibitem{Brunet_2013}
S.~Casale-Brunet, M.~Mattavelli, J.~Janneck, Turnus: A design exploration
  framework for dataflow system design, in: 2013 IEEE International Symposium
  on Circuits and Systems (ISCAS), 2013, pp. 654--654.
\newblock \href {http://dx.doi.org/10.1109/ISCAS.2013.6571927}
  {\path{doi:10.1109/ISCAS.2013.6571927}}.

\bibitem{Bezati_2013}
E.~Bezati, M.~Mattavelli, J.~Janneck, High-level synthesis of dataflow programs
  for signal processing systems, in: 2013 8th International Symposium on Image
  and Signal Processing and Analysis (ISPA), 2013, pp. 750--754.
\newblock \href {http://dx.doi.org/10.1109/ISPA.2013.6703837}
  {\path{doi:10.1109/ISPA.2013.6703837}}.

\bibitem{Serot_2013}
J.~S{\'e}rot, F.~Berry, S.~Ahmed, CAPH: A Language for Implementing
  Stream-Processing Applications on FPGAs, Springer New York, 2013, pp.
  201--224.
\newblock \href {http://dx.doi.org/10.1007/978-1-4614-1362-2_9}
  {\path{doi:10.1007/978-1-4614-1362-2_9}}.

\bibitem{Serot_2016}
J.~S{\'e}rot, F.~Berry, C.~Bourrasset, High-level dataflow programming for
  real-time image processing on smart cameras, {Journal of Real-Time Image
  Processing} 12~(4) (2016) 635--647.
\newblock \href {http://dx.doi.org/10.1007/s11554-014-0462-6}
  {\path{doi:10.1007/s11554-014-0462-6}}.

\bibitem{Serot_2008}
J.~S{\'e}rot, The semantics of a purely functional graph notation system, in:
  Achten, P., Koopman, P.W.M., Moraz{\'a}n, M.T. (eds.) Draft Proceedings of
  the Ninth Symposium on Trends in Functional Programming (TFP), 2008.

\bibitem{Shen_2010}
C.~Shen, W.~Plishker, H.~Wu, S.~S. Bhattacharyya, A lightweight dataflow
  approach for design and implementation of {SDR} systems, in: Proceedings of
  the Wireless Innovation Conference and Product Exposition, 2010, pp.
  640--645.

\bibitem{Zhang_2006}
Y.~Zhang, J.~Roivainen, A.~Mammela, Clock-gating in fpgas: A novel and
  comparative evaluation, in: 9th EUROMICRO Conference on Digital System
  Design: Architectures, Methods and Tools (DSD), 2006, pp. 584--590.
\newblock \href {http://dx.doi.org/10.1109/DSD.2006.32}
  {\path{doi:10.1109/DSD.2006.32}}.

\bibitem{Pedram_1996}
M.~Pedram, Power minimization in ic design: principles and applications, ACM
  Transactions on Design Automation of Electronic Systems 1 (1996) 3--56.
\newblock \href {http://dx.doi.org/10.1145/225871.225877}
  {\path{doi:10.1145/225871.225877}}.

\bibitem{Wu_2000}
Q.~Wu, M.~Pedram, X.~Wu, Clock-gating and its application to low power design
  of sequential circuits, IEEE Transactions on Circuits and Systems I:
  Fundamental Theory and Applications 47~(3) (2000) 415--420.
\newblock \href {http://dx.doi.org/10.1109/81.841927}
  {\path{doi:10.1109/81.841927}}.

\bibitem{Cadence:RTL}
Cadence\textregistered, Using Encounter\textregistered RTL Compiler, Product
  Version 14.1 (July 2014).

\bibitem{Cadence:Genus}
Cadence\textregistered,
  \href{https://www.cadence.com/content/cadence-www/global/en_US/home/tools/digital-design-and-signoff/synthesis/genus-synthesis-solution.html}{Genus
  synthesis solution} (2018).
\newline\urlprefix\url{https://www.cadence.com/content/cadence-www/global/en_US/home/tools/digital-design-and-signoff/synthesis/genus-synthesis-solution.html}

\bibitem{Synopsys:DC}
Synopsys\textregistered,
  \href{https://www.synopsys.com/support/training/rtl-synthesis/design-compiler-rtl-synthesis.html}{Design
  compiler: Rtl synthesis} (2018).
\newline\urlprefix\url{https://www.synopsys.com/support/training/rtl-synthesis/design-compiler-rtl-synthesis.html}

\bibitem{OZBALTAN_2018}
M.~\"{O}zbaltan, N.~Berthier, Exercising symbolic discrete control for
  designing low-power hardware circuits: an application to clock-gating,
  IFAC-PapersOnLine 51~(7) (2018) 120 -- 126, 14th IFAC Workshop on Discrete
  Event Systems (WODES).
\newblock \href {http://dx.doi.org/10.1016/j.ifacol.2018.06.289}
  {\path{doi:10.1016/j.ifacol.2018.06.289}}.

\bibitem{Bezati_2017}
E.~Bezati, S.~Casale-Brunet, M.~Mattavelli, J.~W. Janneck, Clock-gating of
  streaming applications for energy efficient implementations on fpgas, IEEE
  Transactions on Computer-Aided Design of Integrated Circuits and Systems
  36~(4) (2017) 699--703.
\newblock \href {http://dx.doi.org/10.1109/TCAD.2016.2597215}
  {\path{doi:10.1109/TCAD.2016.2597215}}.

\bibitem{Herbert_2007}
S.~Herbert, D.~Marculescu, Analysis of dynamic voltage/frequency scaling in
  chip-multiprocessors, in: Proceedings of the 2007 international symposium on
  Low power electronics and design (ISLPED), 2007, pp. 38--43.
\newblock \href {http://dx.doi.org/10.1145/1283780.1283790}
  {\path{doi:10.1145/1283780.1283790}}.

\bibitem{Eyerman_2011}
S.~Eyerman, L.~Eeckhout, Fine-grained {DVFS} using on-chip regulators, ACM
  Transactions on Architecture and Code Optimization (TACO) 8~(1) (2011) 1--24.
\newblock \href {http://dx.doi.org/10.1145/1952998.1952999}
  {\path{doi:10.1145/1952998.1952999}}.

\bibitem{Arora_2014}
M.~Arora, S.~Manne, Y.~Eckert, I.~Paul, N.~Jayasena, D.~M. Tullsen, A
  comparison of core power gating strategies implemented in modern hardware,
  in: {ACM} International Conference on Measurement and Modeling of Computer
  Systems ({SIGMETRICS}), 2014, pp. 559--560.
\newblock \href {http://dx.doi.org/10.1145/2591971.2592017}
  {\path{doi:10.1145/2591971.2592017}}.

\bibitem{Jeff_2012}
B.~Jeff, Advances in big.little technology for power and energy savings
  improving energy efficiency in high-performance mobile platforms, in: ARM
  White Paper, 2012.

\bibitem{IEEE:UPF}
IEEE Standard for Design and Verification of Low-Power, Energy-Aware Electronic
  Systems, IEEE Standard 1801-2015, UPF-2.0, Unified Power Format 2.0 (2016).

\bibitem{SI2CPFspecification}
Silicon Integration Initiative., Si2 Common Power Format
  Specification\textsuperscript{TM} - Version 2.1 (Dec. 2014).

\bibitem{Gagarski_2016}
K.~Gagarski, M.~Petrov, M.~Moiseev, I.~Klotchkov, Power specification,
  simulation and verification of systemc designs, in: 2016 IEEE East-West
  Design Test Symposium (EWDTS), 2016, pp. 1--4.
\newblock \href {http://dx.doi.org/10.1109/EWDTS.2016.7807731}
  {\path{doi:10.1109/EWDTS.2016.7807731}}.

\bibitem{Qamar_2016}
A.~Qamar, F.~B. Muslim, J.~Iqbal, L.~Lavagno, Lp-hls: Automatic power-intent
  generation for high-level synthesis based hardware implementation flow,
  Microprocessors and Microsystems 50 (2017) 26 -- 38.
\newblock \href {http://dx.doi.org/10.1016/j.micpro.2017.02.002}
  {\path{doi:10.1016/j.micpro.2017.02.002}}.

\bibitem{Macko_2018}
D.~Macko, Contribution to automated generating of system power-management
  specification, in: 2018 IEEE 21st International Symposium on Design and
  Diagnostics of Electronic Circuits Systems (DDECS), 2018, pp. 27--32.
\newblock \href {http://dx.doi.org/10.1109/DDECS.2018.00012}
  {\path{doi:10.1109/DDECS.2018.00012}}.

\bibitem{Carta_2006}
S.~Carta, D.~Pani, L.~Raffo, Reconfigurable coprocessor for multimedia
  application domain, Journal of VLSI signal processing systems for signal,
  image and video technology 44~(1) (2006) 135--152.
\newblock \href {http://dx.doi.org/10.1007/s11265-006-7512-7}
  {\path{doi:10.1007/s11265-006-7512-7}}.

\bibitem{Kumar_2006}
V.~Kumar, J.~Lach, Highly flexible multimode digital signal processing systems
  using adaptable components and controllers, EURASIP Journal on Applied Signal
  Processing 2006 (2006) 73--73.
\newblock \href {http://dx.doi.org/10.1155/ASP/2006/79595}
  {\path{doi:10.1155/ASP/2006/79595}}.

\bibitem{Souza_2005}
C.~C.~d. Souza, A.~M. Lima, G.~Araujo, N.~B. Moreano, The datapath merging
  problem in reconfigurable systems: Complexity, dual bounds and heuristic
  evaluation, Journal of Experimental Algorithmics 10 (2005) 2.2–es.
\newblock \href {http://dx.doi.org/10.1145/1064546.1180613}
  {\path{doi:10.1145/1064546.1180613}}.

\bibitem{Moreano_2002}
N.~Moreano, G.~Araujo, Z.~Huang, S.~Malik, Datapath merging and interconnection
  sharing for reconfigurable architectures, in: 15th International Symposium on
  System Synthesis, 2002, pp. 38--43.
\newblock \href {http://dx.doi.org/10.1145/581199.581210}
  {\path{doi:10.1145/581199.581210}}.

\bibitem{Synflow}
{Synflow SAS}, \href{http://www.synflow.com/}{Synflow ide} (2018).
\newline\urlprefix\url{http://www.synflow.com/}

\bibitem{Fanni_2015}
T.~Fanni, C.~Sau, L.~Raffo, F.~Palumbo, Automated power gating methodology for
  dataflow-based reconfigurable systems, in: Proceedings of the 12th ACM
  International Conference on Computing Frontiers (CF), 2015, pp. 61:1--61:6.
\newblock \href {http://dx.doi.org/10.1145/2742854.2747285}
  {\path{doi:10.1145/2742854.2747285}}.

\bibitem{lowPowGuide}
{P}ower {F}orward {I}nitiative.,
  \href{https://projects.si2.org/events_dir/2009/PowerForward/LowPowerGuide09232009/pfi_lpg_chapters/lpg_sect1_06052009.pdf}{A
  {P}ractical {G}uide to {L}ow {P}ower {D}esign} (june 2009).
\newline\urlprefix\url{https://projects.si2.org/events_dir/2009/PowerForward/LowPowerGuide09232009/pfi_lpg_chapters/lpg_sect1_06052009.pdf}

\bibitem{Fanni_2016}
T.~Fanni, C.~Sau, P.~Meloni, L.~Raffo, F.~Palumbo, Power and clock gating
  modelling in coarse grained reconfigurable systems, in: Proceedings of the
  ACM International Conference on Computing Frontiers (CF), 2016, pp. 384--391.
\newblock \href {http://dx.doi.org/10.1145/2903150.2911713}
  {\path{doi:10.1145/2903150.2911713}}.

\bibitem{Palumbo_2015}
F.~Palumbo, T.~Fanni, C.~Sau, P.~Meloni, L.~Raffo, Modelling and automated
  implementation of optimal power saving strategies in coarse-grained
  reconfigurable architectures, Journal of Electrical and Computer Engineering
  (2016) 27\href {http://dx.doi.org/10.1155/2016/4237350}
  {\path{doi:10.1155/2016/4237350}}.

\bibitem{Xilinx:axi}
Xilinx,
  \href{https://www.xilinx.com/support/documentation/ip_documentation/axi_ref_guide/latest/ug1037-vivado-axi-reference-guide.pdf}{Vivado
  Design Suite --- AXI Reference Guide --- UG1037 (v4.0)} (July 2017).
\newline\urlprefix\url{https://www.xilinx.com/support/documentation/ip_documentation/axi_ref_guide/latest/ug1037-vivado-axi-reference-guide.pdf}

\bibitem{Buss2004}
S.~R. Buss, J.-S. Kim, Selectively damped least squares for inverse kinematics,
  Journal of Graphics Tools 10 (2004) 37--49.
\newblock \href {http://dx.doi.org/10.1080/2151237X.2005.10129202}
  {\path{doi:10.1080/2151237X.2005.10129202}}.

\bibitem{Buss2009}
S.~R. Buss, Introduction to inverse kinematics with jacobian transpose,
  pseudoinverse and damped least squares methods, unpublished. \\ URL
  \url{https://www.math.ucsd.edu/~sbuss/ResearchWeb/ikmethods/iksurvey.pdf}
  (2009).

\bibitem{FanniSRSTP19}
L.~Fanni, L.~Suriano, C.~Rubattu, P.~S{\'{a}}nchez, E.~de~la Torre, F.~Palumbo,
  \href{http://ceur-ws.org/Vol-2457/11.pdf}{A dataflow implementation of
  inverse kinematics on reconfigurable heterogeneous mpsoc}, in: Proceedings of
  the Cyber-Physical Systems PhD Workshop 2019, an event held within the CPS
  Summer School "Designing Cyber-Physical Systems - From concepts to
  implementation", 2019, pp. 107--118.
\newline\urlprefix\url{http://ceur-ws.org/Vol-2457/11.pdf}

\bibitem{PREESM}
M.~Pelcat, K.~Desnos, J.~Heulot, C.~Guy, J.-F. Nezan, S.~Aridhi, Preesm: A
  dataflow-based rapid prototyping framework for simplifying multicore dsp
  programming, in: 2014 6th European Embedded Design in Education and Research
  Conference (EDERC), 2014, pp. 36--40.
\newblock \href {http://dx.doi.org/10.1109/EDERC.2014.6924354}
  {\path{doi:10.1109/EDERC.2014.6924354}}.

\bibitem{Bezati_2013_ESL}
E.~{Bezati}, S.~C. {Brunet}, M.~{Mattavelli}, J.~W. {Janneck}, Synthesis and
  optimization of high-level stream programs, in: Proceedings of the 2013
  Electronic System Level Synthesis Conference (ESLsyn), 2013, pp. 1--6.

\bibitem{bambu}
Bambu.
\newblock \href{https://panda.dei.polimi.it/?page_id=31}{[link]}.
\newline\urlprefix\url{https://panda.dei.polimi.it/?page_id=31}

\bibitem{vivado_hls}
Xilinx,
  \href{www.xilinx.com/products/design-tools/vivado/integration/esl-design}{{Xilinx
  Vivado High-Level Synthesis}}.
\newline\urlprefix\url{www.xilinx.com/products/design-tools/vivado/integration/esl-design}

\bibitem{intel_sdk}
Intel,
  \href{https://www.altera.com/products/design-software/embedded-software-developers/opencl/}{{Intel
  FPGA SDK for OpenCL}}.
\newline\urlprefix\url{https://www.altera.com/products/design-software/embedded-software-developers/opencl/}

\bibitem{cadence_stratus}
Cadence,
  \href{www.cadence.com/content/cadence-www/global/en\_US/home/tools/digital-design-and-signoff/synthesis/stratus-high-level-synthesis}{Stratus
  high-level synthesis}.
\newline\urlprefix\url{www.cadence.com/content/cadence-www/global/en\_US/home/tools/digital-design-and-signoff/synthesis/stratus-high-level-synthesis}

\bibitem{Nane2016}
R.~Nane, V.-M. Sima, C.~Pilato, J.~Choi, B.~Fort, A.~Canis, Y.~T. Chen,
  H.~Hsiao, S.~Brown, F.~Ferrandi, J.~Anderson, K.~Bertels, A survey and
  evaluation of fpga high-level synthesis tools, Trans. Comp.-Aided Des. Integ.
  Cir. Sys. 35~(10) (2016) 1591–1604.
\newblock \href {http://dx.doi.org/10.1109/TCAD.2015.2513673}
  {\path{doi:10.1109/TCAD.2015.2513673}}.

\bibitem{Rubattu_2018_Embedded}
C.~{Rubattu}, F.~{Palumbo}, C.~{Sau}, R.~{Salvador}, J.~{Sérot}, K.~{Desnos},
  L.~{Raffo}, M.~{Pelcat}, Dataflow-functional high-level synthesis for
  coarse-grained reconfigurable accelerators, IEEE Embedded Systems Letters
  11~(3) (2019) 69--72.
\newblock \href {http://dx.doi.org/10.1109/LES.2018.2882989}
  {\path{doi:10.1109/LES.2018.2882989}}.

\bibitem{Bhattacharyya_2011}
S.~Bhattacharyya, J.~Eker, J.~Janneck, C.~Lucarz, M.~Mattavelli, M.~Raulet,
  \href{http://dx.doi.org/10.1007/s11265-009-0399-3}{Overview of the mpeg
  reconfigurable video coding framework}, J. Signal Process. Syst. 63~(2)
  (2011) 251--263.
\newblock \href {http://dx.doi.org/10.1007/s11265-009-0399-3}
  {\path{doi:10.1007/s11265-009-0399-3}}.
\newline\urlprefix\url{http://dx.doi.org/10.1007/s11265-009-0399-3}

\bibitem{Heulot2014}
J.~Heulot, M.~Pelcat, K.~Desnos, J.~F. Nezan, S.~Aridhi, {SPIDER}: {A}
  {S}ynchronous {P}arameterized and {I}nterfaced {D}ataflow-based {RTOS} for
  multicore {DSPS}, in: 2014 6th European Embedded Design in Education and
  Research Conference (EDERC), 2014, pp. 167--171.
\newblock \href {http://dx.doi.org/10.1109/EDERC.2014.6924381}
  {\path{doi:10.1109/EDERC.2014.6924381}}.

\bibitem{Rubattu_2018_CPSWS}
C.~{Rubattu}, \href{http://ceur-ws.org/Vol-2208/5.pdf}{Dataflow-based
  adaptation framework with coarse-grained reconfigurable accelerators}, in:
  Proceedings of the Cyber-Physical Systems PhD and Postdoc Workshop 2018, an
  event held within the CPS Summer School "Designing Cyber-Physical Systems -
  From Concepts to Implementation" (CPSSS 2018), 2018.
\newline\urlprefix\url{http://ceur-ws.org/Vol-2208/5.pdf}

\bibitem{FanniT_2019}
T.~{Fanni}, D.~{Madronal}, C.~{Rubattu}, C.~{Sau}, F.~{Palumbo}, E.~{Juarez},
  M.~{Pelcat}, C.~{Sanz}, L.~{Raffo}, Run-time performance monitoring of
  heterogenous hw/sw platforms using papi, in: Sixth International Workshop on
  FPGAs for Software Programmers (FSP Workshop), 2019, pp. 1--10.

\bibitem{papi}
PAPI, \href{http://icl.utk.edu/papi/}{Performance {API}} (2019).
\newline\urlprefix\url{http://icl.utk.edu/papi/}

\bibitem{Madronal_2019_Access}
D.~Madro{\~n}al, F.~{Arrestier}, J.~{Sancho}, A.~{Morvan}, R.~{Lazcano},
  K.~{Desnos}, R.~{Salvador}, D.~{Menard}, E.~{Juarez}, C.~{Sanz}, Papify:
  Automatic instrumentation and monitoring of dynamic dataflow applications
  based on papi, IEEE Access 7 (2019) 111801--111812.
\newblock \href {http://dx.doi.org/10.1109/ACCESS.2019.2934223}
  {\path{doi:10.1109/ACCESS.2019.2934223}}.

\bibitem{Rodriguez_2018}
A.~Rodr\'{i}guez, J.~Valverde, J.~Portilla, A.~Otero, T.~Riesgo, E.~de~la
  Torre, Fpga-based high-performance embedded systems for adaptive edge
  computing in cyber-physical systems: The artico3 framework, Sensors 18~(6).
\newblock \href {http://dx.doi.org/10.3390/s18061877}
  {\path{doi:10.3390/s18061877}}.

\bibitem{Fanni_2018}
T.~{Fanni}, A.~{Rodr\'{i}guez}, C.~{Sau}, L.~{Suriano}, F.~{Palumbo},
  L.~{Raffo}, E.~{de la Torre}, Multi-grain reconfiguration for advanced
  adaptivity in cyber-physical systems, in: 2018 International Conference on
  ReConFigurable Computing and FPGAs (ReConFig), 2018, pp. 1--8.
\newblock \href {http://dx.doi.org/10.1109/RECONFIG.2018.8641705}
  {\path{doi:10.1109/RECONFIG.2018.8641705}}.

\bibitem{Sau_2017}
C.~Sau, F.~Palumbo, M.~Pelcat, J.~Heulot, E.~Nogues, D.~Menard, P.~Meloni,
  L.~Raffo, Challenging the best hevc fractional pixel fpga interpolators with
  reconfigurable and multifrequency approximate computing, IEEE Embedded
  Systems Letters 9~(3) (2017) 65--68.
\newblock \href {http://dx.doi.org/10.1109/LES.2017.2703585}
  {\path{doi:10.1109/LES.2017.2703585}}.

\bibitem{Palumbo:2019CF}
F.~Palumbo, T.~Fanni, C.~Sau, L.~Pulina, L.~Raffo, M.~Masin, E.~Shindin, P.~S.
  de~Rojas, K.~Desnos, M.~Pelcat, A.~Rodríguez, E.~Juárez, F.~Regazzoni,
  G.~Meloni, K.~Zedda, H.~Myrhaug, L.~Kaliciak, J.~Andriaanse,
  J.~de~Olivieria~Filho, P.~Muñoz, A.~Toffetti, Cerbero: Cross-layer
  model-based framework for multi-objective design of reconfigurable systems in
  uncertain hybrid environments, in: Proceedings of the 16th ACM International
  Conference on Computing Frontiers (CF), ACM, 2019, pp. 320--325.
\newblock \href {http://dx.doi.org/10.1145/3310273.3323436}
  {\path{doi:10.1145/3310273.3323436}}.

\bibitem{zaid_2019}
Z.~Al{-}Ars, T.~Basten, A.~de~Beer, M.~Geilen, D.~Goswami,
  P.~J{\"{a}}{\"{a}}skel{\"{a}}inen, J.~Kadlec, M.~M. de~Alejandro, F.~Palumbo,
  G.~Peeren, L.~Pomante, F.~van~der Linden, J.~Saarinen, T.~S{\"{a}}ntti,
  C.~Sau, M.~K. Zedda, The fitoptivis {ECSEL} project: highly efficient
  distributed embedded image/video processing in cyber-physical systems, in:
  Proceedings of the 16th {ACM} International Conference on Computing Frontiers
  ({CF}), 2019, pp. 333--338.
\newblock \href {http://dx.doi.org/10.1145/3310273.3323437}
  {\path{doi:10.1145/3310273.3323437}}.

\bibitem{Masin_2017}
M.~Masin, F.~Palumbo, H.~Myrhaug, J.~A. de~Oliveira~Filho, M.~Pastena,
  M.~Pelcat, L.~Raffo, F.~Regazzoni, A.~A. Sanchez, A.~Toffetti, E.~de~la
  Torre, K.~Zedda, Cross-layer design of reconfigurable cyber-physical systems,
  in: Design, Automation Test in Europe Conference Exhibition (DATE), 2017,
  2017, pp. 740--745.
\newblock \href {http://dx.doi.org/10.23919/DATE.2017.7927088}
  {\path{doi:10.23919/DATE.2017.7927088}}.

\end{thebibliography}

\end{document}